\documentclass[aps,prx,twocolumn,showpacs]{revtex4-1}
\usepackage{bm}
\usepackage{mathrsfs}
\usepackage{amsmath}
\usepackage{amssymb}
\usepackage{graphicx}
\usepackage{amsfonts}
\usepackage{amsthm}
\usepackage{color}
\usepackage{dcolumn}
\usepackage{txfonts}
\usepackage[caption=false]{subfig}
\usepackage[T1]{fontenc}
\usepackage{multirow}

\begin{document}

\title{Nonlinear optical realization of non-integrable phases accompanying quantum phase transitions}
\author{Chon-Fai Kam}
\email{dubussygauss@gmail.com}
\affiliation{Department of Physics, University at Buffalo, SUNY, Buffalo, New York 14260-1500, USA}

\begin{abstract}
In this work, we propose an experimentally feasible nonlinear optical realization of a type of non-integrable phase found in interacting quantum systems at quantum phase transitions. We show that an exotic term in the dynamical equation governs the nonlinear polarization of the optical field along an anisotropic low-birefringence fiber with tetragonal symmetry. Intriguingly, by adiabatically tuning nonlinear susceptibilities along the fibers, the Stokes vector on the Poincar\'e sphere accumulates a non-integrable phase called the Hannay angle, which shares the same geometric gauge structure as that associated with quantum phase transitions. Experimental realization via adiabatically depositing nano-crystals along the fibers is discussed.
\end{abstract}

\maketitle

\section{Introduction}
Non-integrable phase factors \cite{wu1975concept} play a fundamental role in the global formulation of gauge fields. In gauge theory, the base space is represented as a smooth manifold, while wave functions are locally defined as sections of a fiber bundle over this manifold. The gauge potential acts as a connection that glues these local sections, adhering to a symmetry governed by a Lie group—forming a common framework in high-energy physics. Non-trivial topology can be identified algebraically through the computation of homology groups \cite{atiyah2005geometry}. Alternatively, a wave vector can be transported along a closed loop via a prescribed parallel transport law on the manifold, allowing one to observe discrepancies between the final and initial states. This deviation, manifested as a non-vanishing phase, is known as anholonomy and may correspond to a group element in both Abelian and non-Abelian gauge theory. This differential topology approach is crucial in determining whether a local gauge field can be extended to a global one \cite{steenrod1999topology}.

Interestingly, this framework naturally extends to a much broader context and is ubiquitous in physics. For example, consider a quantum system coupled to an external classical system—effectively modeled as a set of external parameters. A typical case involves a spin in a classical electromagnetic field. As the system undergoes a cyclic adiabatic evolution of these parameters, the wavefunction accumulates a geometric phase factor that cannot be removed by any gauge transformation \cite{berry1984quantal, kam2023coherent}. This highlights the role of adiabatic evolution in probing the underlying gauge field within parameter space. This non-vanishing phase factor, known as the Berry phase, has played a central role in the historical development of quantum Hall effects \cite{simon1983holonomy}. In condensed matter physics, the same geometric principles govern emergent properties in exotic materials, such as topological systems \cite{wen2004quantum} and spin liquids. The universality of geometric phases underscores the elegance and power of this geometric language. Additionally, Berry phases are fundamental phenomena spanning various areas of physics, from molecular systems \cite{bohm2013geometric} and optical waveguides \cite{shapere1989geometric} to quantum entanglement \cite{kam2021berry}. 

Beyond physics, this geometric framework seamlessly extends into various natural and biological systems. For example, in fluid mechanics, microorganisms like bacteria navigate environments where viscous forces dominate inertial forces at low Reynolds numbers. Their movement relies on non-reciprocal motions—such as rotating flagella or coordinated shape changes—to generate propulsion constrained by geometric principles \cite{shapere1989geometry}. Similarly, in biomechanics, the "cat-righting reflex"—the way a cat reorients itself mid-fall—relies on the conservation of angular momentum and a subtle redistribution of body segments \cite{shapere1989gauge}. By twisting its torso in a carefully choreographed sequence, a cat can achieve an effective rotation despite starting from a zero-angular-momentum state. Beyond living systems, geometric effects emerge in unexpected domains. In robotics, autonomous swarms of drones or underwater vehicles use geometric phase concepts to coordinate movements efficiently, mimicking biological swarming behaviors \cite{brambilla2013swarm, hatton2015nonconservativity}. 

In particular, Berry phase factors are accumulated through a gauge-invariant field known as the Berry curvature when a system undergoes cyclic adiabatic evolution. In a finite-dimensional quantum system, the gauge-invariant Berry curvature diverges at energy-level crossing points, a phenomenon that can be understood as a magnetic singularity akin to a monopole in parameter space \cite{berry1984quantal}. In many-body systems, quantum phase transitions are marked by the crossing of the lowest energy levels. Thus, at quantum phase transitions, the Berry curvature exhibits point-like singularities, and the associated geometric phase becomes a topological invariant—an effect well-documented in studies of topological phase transitions and critical points within parameter space \cite{sachdev1999quantum}. However, in excited-state quantum phase transitions \cite{caprio2008excited}, where higher energy levels intersect, the behavior of Berry curvature can be more nuanced, depending on the nature of the level crossings and the structure of the Hamiltonian.

In certain cases of quantum phase transitions, not only does the ground state of the system transition from one symmetry phase to another, but the entire spectral symmetry generated by the dynamical group changes across the phase boundary \cite{cejnar2010quantum}. This phenomenon differs from conventional ground-state quantum phase transitions. Consequently, the Berry curvature is expected to exhibit singular behaviors distinct from the point-like singularity observed in excited-state quantum phase transitions and may possess richer geometric structures. A prime example can be found in cold atom systems such as Bose-Einstein condensates, where the Berry curvature diverges at a bifurcation surface during the excited state quantum phase transition from Rabi oscillations to macroscopic self-trapping, where the singularity in the gauge field exhibits the same geometric structure as the cosmological singularity found in a de Sitter universe \cite{kam20172+}. Due to experimental difficulty in precisely controlling Bose-Einstein condensates near quantum phase transitions, the realization of the exotic geometric structure of Berry curvature singularities in interacting quantum systems remains elusive. Nevertheless, given that the geometric phase was first realized in classical optical waveguides \cite{tomita1986observation}, it would be intriguing to first experimentally realize this novel gauge field singularities within optical systems. Such a demonstration could provide insights into the interplay between topology and wave dynamics in these settings. Benefit from the well-developed techniques in the field of nonlinear guided wave optics, we will demonstrate in the subsequent sections, that the Berry curvature singularity in interacting quantum systems may be realized by the geometric phase shift of nonlinear polarization in anisotropic fibers, when the nonlinear susceptibilities undergo cyclic adiabatic variation.

\section{Geometric phases in optical systems}\label{5.2}
The concept of geometric phases in optical systems dates back to the mid-1950s when Pancharatnam published his research on phase shifts in polarized light \cite{pancharatnam1955achromatic, pancharatnam1956generalized}. As a historical fact, his work preceded Berry's work on quantum adiabatic phases by almost three decades, but was forgotten in the following decades until its rediscovery by Ramaseshan and Nityananda in 1986 \cite{ramaseshan1986interference}. Pancharatnam's idea is simple and elegant --- he found a way to compare the phases between distinct and non-orthogonal polarized states on the Poincare sphere. For monochromatic light propagating along the $z$-direction with its electric field lying in the $xy$-plane, the state of polarization may be described by two complex wave amplitudes of the electric field, $E_x$ and $E_y$, which satisfy $|E_x|^2+|E_y|^2=1$ and may be used to form a two-component spinor \cite{berry1987adiabatic}
\begin{equation}
|\psi\rangle=
\left( 
\begin{array}{ccc}
\psi_+  \\
\psi_- 
\end{array} \right),
\psi_\pm \equiv \frac{1}{\sqrt{2}}(E_x\pm iE_y),
\end{equation}
where each spinor is an eigenvector of some $2\times 2$ Hermitian polarization matrix of the form
\begin{equation}
H(\textit{\textbf{r}})=\textit{\textbf{r}}\cdot \boldsymbol{\sigma}=\frac{1}{2}
\left[ \begin{array}{ccc}
\cos\theta & \sin\theta e^{-i\phi}  \\
\sin\theta e^{i\phi} & -\cos\theta 
\end{array} \right].
\end{equation}
The Poincare sphere is a sphere with spherical coordinates $\theta$ and $\phi$. Each position on the sphere, $\textit{\textbf{r}}_A$, defines a polarization state $|\psi(\textit{\textbf{r}}_A)\rangle$, which is an eigenvector of $H$ and may simply be denoted as $|A\rangle$. Pancharatnam suggested that the phase difference between two distinct polarization states $|A\rangle$ and $|B\rangle$ may be defined as the phase $\arg \langle A|B\rangle$ of their scalar product. Hence, $|A\rangle$ and $|B\rangle$ are in phase when the scalar product $\langle A|B\rangle$ is real and positive. Pancharatnam considered the case when a state $|A\rangle$ passes through a sequence of polarization analyzers, such as wave plates or retarders, so that state $|A\rangle$ becomes state $|B\rangle$ after the first analyzer, $|B\rangle$ becomes $|C\rangle$ after the second analyzer, and $|C\rangle$ returns to state $|A'\rangle$ after the third analyzer, which corresponds to the same point on the Poincare sphere. He showed that even if $|A\rangle$ is in phase with $|B\rangle$, and $|B\rangle$ is in phase with $|C\rangle$, $|C\rangle$ still need not be in phase with $|A\rangle$ due to the path traversed on the Poincare sphere \cite{berry1994pancharatnam}. When the state returns to the same point on the Poincare sphere, the phase difference between the initial and final states is given by $\langle A|A'\rangle=\exp(-i\Omega_{ABC}/2)$, where $\Omega_{ABC}$ is the solid angle of the geodesic triangle $ABC$. In fact, Pancharatnam's phase is identical to Berry's adiabatic geometric phase of a spin-1/2 particle in a slowly varying magnetic field.

As a historical coincidence, the first experimental observation of Berry's phase was also performed in optical systems. In 1986, Chiao and Wu recognized that Berry's phase may be realized in optical waveguides which exploit spin-orbit interactions of light in which the pseudo-spin of a photon $\textit{\textbf{s}}$ follows the direction of momentum $\textit{\textbf{k}}$ --- in other words, the helicity $\textit{\textbf{s}}\cdot\textit{\textbf{k}}$ is an adiabatic invariant for slow variations of the local direction of light down a helical waveguide \cite{chiao1986manifestations}. They argued that the evolution of the spin of a photon propagating along a helically wound single mode circular waveguide is governed by a Hamiltonian $H(\textit{\textbf{k}})=H_0+\textit{\textbf{s}}\cdot\textit{\textbf{k}}$, which has a form similar to that of a particle with spin $\textit{\textbf{s}}$ in a slowly varying magnetic field, where the only difference is the replacement of the parameter space by momentum space. In their proposal, linear birefringence in the media and shape variation in the core cross-section are both neglected, as they may cause coupling between the states of opposite helicity. Chiao and Wu's proposal was later verified by Tomita and Chiao in the same year \cite{tomita1986observation}. Their experiment used a 180 cm long helically coiled single mode fiber made of linear media, which had a core diameter of 2.6 $\mu$m. Their experiment showed that the plane of polarization of light becomes rotated by an angle which is equal to Berry's phase, provided that the input and output ends of the fiber are kept aligned. 

Within the same year Chiao and Wu's work was published, Berry proposed another way to realize adiabatic geometric phases in optical systems \cite{berry1986adiabatic}, where he suggested the usage of anisotropic dielectric waveguide made of non-absorbing linear media which exhibit both uniaxial birefringence and magneto-optic effects, for which the displacement and electric field are related by $\textit{\textbf{D}}=\epsilon_0\overset\leftrightarrow{\epsilon}\textit{\textbf{E}}$, or equivalently, $\textit{\textbf{E}}=\overset\leftrightarrow{\eta}\textit{\textbf{D}}/\epsilon_0$, where $\overset\leftrightarrow{\epsilon}$ is the dielectric tensor, and $\overset\leftrightarrow{\eta}$ is the inverse dielectric tensor. If the electric field is guided along the $z$-direction, the inverse dielectric tensor in the plane perpendicular to the direction of propagation is a $2 \times 2$ Hermitian matrix described by \cite{landau2013electrodynamics}
\begin{equation}\label{InverseDielectricTensor}
\overset\leftrightarrow{\eta}_\perp=
\left( \begin{array}{ccc}
1/n_o^2+b_x^2 & b_xb_y+ig_z  \\
b_xb_y-ig_z & 1/n_o^2+b_y^2 \end{array} \right),
\end{equation}
where $n_o$ is the ordinary refractive index for the uniaxial crystal, the terms $b_x$ and $b_y$ are caused by the electro-optical Kerr effect, where a birefringence proportional to the square of the applied electric field is induced: $\delta\epsilon_{ij}=\chi_{ijkl}^{(3)}E_kE_l$, and the term $g_z$ is due to linear magneto-optical effects, e.g. the optical Faraday effect, where the polarization of light is rotated by an angle which is proportional to the magnetization: $\delta\epsilon_{ij}=\chi_{ijk}^{(2)}H_k$. Prior to Tomita and Chiao's experiment in the 1980s, Berry's proposal had not yet been realized experimentally --- magneto-optic effects are small in media available at the time, which would have made magnetic field requirements for experiments too high. Interestingly, the inverse dielectric tensor $\overset\leftrightarrow{\eta}_\perp$ has a form similar to the Hamiltonian of a spin-1/2 particle in a slowly varying magnetic field. Motivated by this similarity, Berry recognized that the eigenstates of $\overset\leftrightarrow{\eta}_\perp$ also accumulate a non-vanishing geometric phase, which may be induced by a slow rotation of the axis of birefringence and the axis of gyrotropy \cite{berry1990budden}. If the optical axes defined by birefringence and gyrotropy have spherical angles $\theta$ and $\phi$ relative to the $z$-axis, so that $b_x=b\sin\theta\cos\phi$, $b_y=b\sin\theta\sin\phi$ and $g_z=g\cos\theta$, the Berry phase for keeping $\theta$ constant and increasing $\phi$ from $0$ to $2\pi$ has the form \cite{berry1986adiabatic}
\begin{equation}
\gamma(\theta) = \pm 2\pi\left(1-\frac{2\cos\theta}{\sqrt{\sigma^2\sin^4\theta+4\cos^2\theta}}\right),
\end{equation}
where $\sigma\equiv b^2/g$. In principle, the terms $b_x$ and $b_y$ in the inverse dielectric tensor $\overset\leftrightarrow{\eta}_\perp$ may also be induced by the Voigt effect \cite{voigt1908magneto} in gases, or the Cotton-Mouton effect \cite{cotton1905cr} in solids and liquids, both of which induce a birefringence quadratic in the magnetization: $\delta\epsilon_{ij}=\chi_{ijkl}^{(3)}H_kH_l$. However, the Cotton-Mouton and Voigt effects are in general too small to have significant effects on geometric phases \cite{haigh2015magneto}. For example, the Voigt rotation angle in a dense cloud of cold $^7$Li atoms is measured to be 0.2$^{\circ}$ at a constant magnetic field of 12.6 G at an initial polarization angle of 45$^{\circ}$ to the magnetic field \cite{franke2001magneto}. Within the same paper \cite{berry1986adiabatic}, Berry also mentioned the possibility of using chiral liquid crystals, which consist of employing a layered arrangement of rod-shaped molecules as the anisotropic media to realize the Hermitian matrix $\overset\leftrightarrow{\eta}_\perp$. Berry's proposal has been implemented recently in photo-aligned nematic liquid crystals by Slussarenko \textit{et al}. \cite{slussarenko2016guiding}, in which the optical axis is purely planar, and undergoes a sinusoidal modulation along the waveguide axis, described quantitatively by a rotation angle $\theta(x,z)\equiv \sin(2\pi z/\Lambda)\Gamma(x)$, where $\Lambda\equiv \lambda/\Delta n$ and $\Gamma(x)$ is the transverse distribution of optical axis orientation. Here, $\lambda$ is the wavelength of the continuous-wave input Gaussian beam and $\Delta n\equiv n_e-n_o$ is the difference between the ordinary and extraordinary refractive indices. In the paraxial limit where the longitudinal field component is neglected, and in low-birefringent uniaxial dielectrics, Slussarenko \textit{et al}. demonstrated that the slowly varying envelope of the circularly polarized electric field in the rotated frame may be described by \cite{slussarenko2016guiding}
\begin{equation}\label{OpticalLinearSchrodingerEq}
i\frac{\partial A}{\partial z}=-\frac{1}{2\bar{n}k_0}\frac{\partial^2A}{\partial x^2}\pm\frac{\pi}{\Lambda}\Gamma(x) A,
\end{equation}
where $\bar{n}\equiv (n_o+n_e)/2$ is the average refractive index, and the plus and minus signs correspond to left and right circularly polarized envelopes, respectively. Eq.\:\eqref{OpticalLinearSchrodingerEq} is equivalent to the one-dimensional Schr\"{o}dinger equation for a particle moving in a potential $V(x)=\pm\pi\Gamma(x)/\Lambda$; therefore, depending on the sign of $\Gamma(x)$, either the left or right circularly polarized envelope of the electric field will be confined in the synthetic optical potential. For a purely circular polarization at the input, Slussarenko \textit{et al}. showed that the Pancharatnam-Berry phase between the two states corresponding to the two distinct orientations of the optical axis, i.e., $\theta_1$ and $\theta_2$, is given by \cite{slussarenko2016guiding}
\begin{equation}
\Delta\varphi(\theta_1,\theta_2)=\arg\left[\cos^2\left(\frac{\delta}{2}\right)+\sin^2\left(\frac{\delta}{2}\right)e^{2i(\theta_2-\theta_1)}\right],
\end{equation}
where $\delta\equiv 2\pi\Delta n z/\lambda$ is the phase retardation between the ordinary and extraordinary eigenfields. Recently, the geometric phase shifter discussed above has been shown to be extendable to any pair of orthogonal polarization states --- linear, circular, or elliptical --- in experimental realizations in which a metasurface consisting of a single-layer array of amorphous silicon posts \cite{arbabi2015dielectric} or titanium dioxide pillars  \cite{mueller2017metasurface}. The technique of geometric phase shifting has been applied to low-coherence interference microscopy \cite{roy2002geometric, roy2004low}, and has been shown to minimize errors due to spurious diffraction effects \cite{roy2009white}. In particular, geometric phase shifting based on fast switchable ferroelectric liquid-crystals is demonstrated to be a novel technique for rapid full-field optical coherence tomography \cite{lu2012full}.

In the last decade, the concept of geometric phases has been extended to time-dependent conducting linear media \cite{choi2005quantum, pedrosa2009electromagnetic}, in which the vector potential $\textit{\textbf{A}}_k(\mathbf{r}, t)$ can be written as the superposition of a set of eigenmodes $\textit{\textbf{u}}_k(\mathbf{r})$ and wave amplitudes $q_k(t)$ as $\textit{\textbf{A}}_k(\mathbf{r}, t)=\sum_k\textit{\textbf{u}}_k(\mathbf{r})q_k(t)$ \cite{maamache2010geometric}. The equations of motion for the wave amplitudes $q_k(t)$ and the canonical conjugate variables $p_k(t)$ are described by the Hamiltonian of a generalized harmonic oscillator
\begin{equation}
H=\frac{1}{2}\left(\frac{p_k^2}{\epsilon}+2\Lambda p_kq_k+\epsilon\Omega_k^2q_k^2\right),
\end{equation}
where $\Lambda(t)\equiv \frac{\sigma(t)}{2\epsilon(t)}$, $\Omega_k(t)\equiv \frac{\omega_k}{c_0}[\mu(t)\epsilon(t)]^{-1/2}$, $\omega_k$ is the natural frequency of the mode $k$, and $c_0$ is the velocity of electromagnetic wave inside the time-dependent linear media at $t=0$. Hence, each mode accumulates a non-vanishing geometric phase, provided that the electric permittivity $\epsilon$, magnetic permeability $\mu$, and conductivity $\sigma$ parametrize a cyclic adiabatic process in time and span the parameter space. The non-adiabatic correction to geometric phases in time-dependent linear media has been studied for the specific case that the electric permittivity varies according to $\epsilon(t)=\epsilon_0(1+b_1e^{-b_2t})$ \cite{lakehal2016novel}. 

\section{Geometric phases for anisotropic fibers}\label{5.3}

While the concept of geometric phases was first introduced in closed quantum systems, which involve the unitary evolution of wave functions under the Schrödinger equation in the adiabatic limit, it is interesting to note that the earliest experimental observation of geometric phases occurred in fiber optics—specifically in helically coiled single-mode optical fibers within linear media \cite{tomita1986observation, berry1987interpreting}. In such experiments, the non-integrable phase factor for polarization of light is associated with a monopole at the origin of the parameter space \cite{bliokh2008geometrodynamics}, similar to that of a spin-1/2 in an cyclic and adiabatically varying magnetic field.

Geometric phases in linear fiber optics have been extensively studied over the last few decades, whereas nonlinear effects in optical fibers become significant for optical pulses with high peak powers \cite{agrawal2000nonlinear}. Despite ongoing advances, research on geometric phases in nonlinear fiber optics remains relatively limited. More importantly, the nonlinear wave equations governing the slowly varying wave amplitudes along the fiber axis closely resemble the mean-field description of interacting quantum systems in nonlinear quantum mechanics. This striking similarity positions nonlinear fiber optics as a powerful experimental platform for investigating emergent gauge fields in various interacting quantum systems.

Building on this foundation, we aim to study geometric phases of light propagating in nonlinear optical waveguides using the theoretical framework of nonlinear quantum mechanics. Under different symmetry conditions, the slowly varying wave amplitudes of polarization components obey different sets of coupled-mode equations, which may be expressed in the form of nonlinear Schr\"{o}dinger equations for specific cases (See App. \ref{C} for a derivation). 

Among the different crystal symmetries, tetragonal symmetry—particularly the three crystal classes $4$, $\bar{4}$, and $4/m$—stands out due to its unique properties. Specifically, for continuous-wave (CW) radiation, the coupled-mode equations simplify as the time derivative terms become negligible. Consequently, the evolution of the polarization state in anisotropic fibers with tetragonal symmetry can be described by a nonlinear Schr\"{o}dinger equation of the form
\begin{align}\label{CMEQforTetragonalSymmetry}
i\frac{d u_j}{d z}=&-\frac{\xi_j}{2}u_j-a|u_j|^2u_j-b\left(2|u_k|^2u_j+u_k^2u_j^*\right)\nonumber\\
&\mp c\left(2|u_j|^2u_k+u_j^2u_k^*\right)\mp d|u_k|^2u_k,
\end{align}
where $\xi_j\equiv \beta_{0j}-\beta_{0k}+i\alpha_j$ is a complex parameter, $\alpha_j$ accounts for fiber losses, and the minus and plus signs correspond to $j=x$ and $j=y$ respectively. Here, the relationship between $u_j$ and the slow-varying wave amplitudes $A_j$ is given by $u_x=A_xe^{i\Delta\beta z/2}$ and $u_y=A_ye^{-i\Delta\beta z/2}$, where $\Delta\beta\equiv \beta_{0x}-\beta_{0y}$ is the difference between the two propagation constants along the slow and fast axes. 

In the following, we may neglect the terms proportional to $\Delta\beta$, i.e., assume that $\xi_j=0$, as we are mainly interested in polarization effects in low-birefringent fibers. For fibers with negligible losses, \eqref{CMEQforTetragonalSymmetry} is integrable only when the condition $d=-c$ is fulfilled, which yields a set of nonlinear Schr\"{o}dinger equations that can be written as
\begin{equation}\label{NonlinearSchrodingerEq}
i\frac{du_j}{dz}=\frac{\partial}{\partial u_j^*}H(u_j,u_j^*),-i\frac{du^*_j}{dz}=\frac{\partial}{\partial u_j}H(u_j,u_j^*),
\end{equation}
where the Hamiltonian $H(u_j,u_j^*)$ is given by
\begin{align}\label{AnisotropicPoincareHamiltonian}
&H=-\frac{1}{2}\left\{c_0(|u_x|^2+|u_y|^2)^2+c_z(|u_x|^2-|u_y|^2)^2\right.\nonumber\\
&\left.+c_x(u_x^*u_y+u_y^*u_x)^2+2c(|u_x|^2-|u_y|^2)(u_x^*u_y+u_y^*u_x)\right\},
\end{align}
where $c_0 = \frac{1}{2}(a+b)$, $c_z = \frac{1}{2}(a-b)$, and $c_x = b$. Equation \eqref{NonlinearSchrodingerEq} has the same form as the nonlinear Schr\"{o}dinger equation in Weinberg's formulation of nonlinear quantum mechanics (see App.\:C for a discussion). 

Notice that the last term in Eq.\:\eqref{AnisotropicPoincareHamiltonian} is an exotic term that exists only in materials with tetragonal symmetry, while other terms exist in other symmetry classes as well. This term leads to a bifurcation in the classical polarization dynamics and results in a nonzero, non-integrable phase when the nonlinear susceptibilities vary adiabatically.

In particular, the Hamiltonian \eqref{AnisotropicPoincareHamiltonian} is a homogeneous function of $u_x$ and $u_y$ of degree 2, which satisfies $H(z u_x,zu_y)=z^2H(u_x,u_y)$ for any complex number $z$. Here, $u_j$ and $u_j^*$ are regarded as independent variables. The evolution of the polarization state can also be studied using the Poincare sphere representation, which is described by the rotation of the Stokes vector on the Poincare sphere \cite{born2013principles}. Here, the Stokes parameters are defined by 
\begin{align}\label{StokesParameters}
S_0&\equiv |u_x|^2+|u_y|^2,\:\;S_z\equiv |u_x|^2-|u_y|^2,\nonumber\\
S_x&\equiv u_x^*u_y+u_y^*u_x,\:\;S_y\equiv -i(u_x^*u_y-u_y^*u_x),
\end{align}
so that the Hamiltonian \eqref{AnisotropicPoincareHamiltonian} may be expressed as
\begin{equation}\label{FinalHamiltonian}
H=-\frac{1}{2}\left(c_0S_0^2+c_zS_z^2+2cS_zS_x+c_xS_x^2\right).
\end{equation}
After a straightforward calculation, \eqref{CMEQforTetragonalSymmetry} with fixed parameters may also be expressed in terms of the Stokes parameters as
\begin{subequations}
\begin{align}\label{AnisotropicSpinEquations1}
\dot{S}_x&=2S_y(cS_x+c_zS_z),\\ 
\dot{S}_y&=2S_z(c_xS_x+cS_z)-2S_x(cS_x+c_zS_z),\\
\dot{S}_z&=-2S_y(c_xS_x+cS_z),\label{AnisotropicSpinEquations3}
\end{align}
\end{subequations}
where $\dot{S}_j$ denotes the derivative of $S_j$ with respect to $z$. For fixed parameters, there are two integrals of motion for \eqref{AnisotropicSpinEquations1} -- \eqref{AnisotropicSpinEquations3}, which are
\begin{subequations}
\begin{align}
S_x^2+S_y^2+S_z^2&=S_0^2,\\
c_zS_z^2+2cS_zS_x+c_xS_x^2&=-2H-c_0S_0^2. 
\end{align}
\end{subequations}
Hence, \eqref{AnisotropicSpinEquations1} -- \eqref{AnisotropicSpinEquations3} are integrable.

In the following, we focus on the polarization dynamics near the fixed points. Without loss of generality, we may assume that $S_0=1$. As we can see from \eqref{AnisotropicSpinEquations1} -- \eqref{AnisotropicSpinEquations3}, $(S_x,S_y,S_z)=(0,\pm 1,0)$ are two fixed points of the polarization dynamics, which implies that $u_x$ and $u_y$ are equal in amplitude, but have a phase difference of $\pi/2$. Near the positive $y$-axis on the Poincare sphere, \eqref{AnisotropicSpinEquations1} -- \eqref{AnisotropicSpinEquations3} become
\begin{equation}\label{GeHarmonicsOscillator}
\dot{S_x}=2cS_x+2c_zS_z,
\dot{S_z}=-2cS_z-2c_x S_x.
\end{equation}
\eqref{GeHarmonicsOscillator} is equivalent to the dynamics of a generalized harmonic oscillator, and is described by the Hamiltonian $H=(\alpha q^2+2\beta pq+\gamma p^2)/2$, where $S_x$ and $S_z$ are identified as the canonical coordinates by the relations $q\equiv S_x$ and $p\equiv S_z$, and the parameters $\alpha$, $\beta$ and $\gamma$ are given by $\alpha\equiv 2c_x$, $\beta\equiv 2c$ and $\gamma\equiv 2c_z$. Similarly, near the negative $y$-axis on the Poincare sphere, \eqref{AnisotropicSpinEquations1} -- \eqref{AnisotropicSpinEquations3} become
\begin{equation}
\dot{S_x}=-2cS_x-2c_zS_z,
\dot{S_z}=2cS_z+2c_x S_x,
\end{equation}
which is described by the Hamiltonian $H=(\alpha q^2+2\beta pq+\gamma p^2)/2$, where $S_x$ and $S_z$ are identified as the canonical coordinates by the relations $q\equiv S_x$ and $p\equiv S_z$, and the parameters $\alpha$, $\beta$ and $\gamma$ are given by $\alpha\equiv -2c_x$, $\beta\equiv -2c$ and $\gamma\equiv -2c_z$. 

The dynamics of a generalized harmonic oscillator is \textit{stable} only when $\alpha\gamma-\beta^2>0$, where $\omega\equiv\sqrt{\alpha\gamma-\beta^2}$ is the frequency of oscillation, and the associated contours for fixed parameters are concentric oblique ellipses in phase space. Hence, $\alpha\gamma-\beta^2=0$ forms a critical surface of bifurcation in parameter space. In contrast, when $\alpha\gamma-\beta^2<0$, or when the exotic term in Eq.\:\eqref{AnisotropicPoincareHamiltonian} dominates, the polarization dynamics will deviate quantitatively from that of cubic symmetry, which will be reflected in the singular behavior of the classical geometric phase. As we can observe from \eqref{StokesParameters}, all the Stokes parameters are invariant under a redefinition of the phase of the slowly varying wave amplitudes $u_j\rightarrow e^{i\lambda}u_j$, and hence the overall phase of $u_j$ does not appear in the equations of motion for the Stokes parameters, \eqref{AnisotropicSpinEquations1} -- \eqref{AnisotropicSpinEquations3}, in the Poincare sphere representation. For a redefinition of the phases of the wave amplitudes $u_j\equiv e^{i\lambda} v_j$, the nonlinear Schr\"{o}dinger equations for the new amplitudes $v_j$ become $i\dot{v}_j-v_j\dot{\lambda}=\partial H(v_j,v_j^*)/\partial v_j^*$, and hence the overall phase $\lambda$ is determined by
\begin{equation}
\dot{\lambda}=i\sum_j v_j^*\dot{v}_j-\sum_j v_j^*\frac{\partial H(v_j,v_j^*)}{\partial v_j^*}=i\sum_j v_j^*\dot{v}_j-2H(v_j,v_j^*),
\end{equation}
where we have used the condition that the Hamiltonian $H(v_j,v_j^*)$ is a homogeneous function of $v_x$ and $v_y$ of degree 2. To be specific, we assign the overall phase $\lambda$ to be the average of the phases of $u_x$ and $u_y$, $\lambda\equiv (\arg u_x+\arg u_y)/2$, so that we may write $v_x\equiv \sqrt{\frac{1+p}{2}}e^{iq}$ and  $v_y\equiv \sqrt{\frac{1-p}{2}}e^{-iq}$ for the polarization states near the positive $y$-axis on the Poincare sphere. A direct calculation yields $\dot{\lambda}=-p\dot{q}+H(p,q)$, where $H(p,q)=c_0+(\alpha q^2+2\beta pq+\gamma p^2)/2$ is the Hamiltonian of the generalized harmonic oscillator described above. 

After integration over a period $T\equiv 2\pi/\omega$, the overall phase satisfies $\lambda(T)=\lambda(0)-2\pi I + \int_0^T H(p,q) dz$, where $I\equiv \frac{1}{2\pi}\oint pdq$ is the classical action for the generalized harmonic oscillator. For fixed parameters, the Hamiltonian of the generalized harmonic oscillator may be expressed as $H(p,q)=c_0+I\omega$, and hence we obtain $\lambda(T)=\lambda(0)+ c_0T$. We now consider the case when the parameters of the generalized harmonic oscillators $\alpha$, $\beta$ and $\gamma$ are adiabatically varied along the fiber. The slowly varying wave amplitudes $u_x$ and $u_y$ will accumulate a geometric phase, namely, the Hannay angle, provided that the parameters $\alpha$, $\beta$ and $\gamma$ are adiabatically varied, and form a closed loop $C$ in parameter space (See App.\:\ref{A} for supporting content). In such a case, we may write the overall phase as $\lambda(T)=\lambda(0)+c_0T+\gamma_H$, where $\gamma_H$ denotes the Hannay angle, which only depends on the shape of the contour in parameter space, and can be written as an integral of an angle 2-form $W$ over an open surface $S$ whose boundary is $C$ 
\begin{equation}
\gamma_H=\int_{\partial S=C} W,
\end{equation}
where the angle 2-form $W$ is expressed as
\begin{equation}\label{Singularity}
W=\frac{\alpha d\beta \wedge d\gamma+\beta d\gamma \wedge d\alpha+\gamma d\alpha\wedge d\beta}{4(\alpha\gamma-\beta^2)^{3/2}}.
\end{equation}
From Eq.\:\eqref{Singularity}, we see that the angle 2-form possesses a \textit{conic singularity} at the critical surface $\alpha\gamma-\beta^2=0$ in parameter space, which reflects the bifurcation of polarization dynamics on the Poincare sphere. 

The above discussion implies that the polarization state of the optical field, represented by the Stokes parameters, experiences a bifurcation in its dynamics when the exotic term in Eq.\:\eqref{AnisotropicPoincareHamiltonian} dominates. The gauge field singularity in Eq.\:\eqref{Singularity} is different from the conventional magnetic field singularity of a monopole, but is the same as the one associated with a quantum phase transition from Rabi oscillations to macroscopic self-trapping in Bose-Einstein condensates in an asymmetric double-well potential \cite{kam20172+}. The only difference lies in that the time $t$ in Bose-Einstein condensate cases now corresponds to the fiber direction $z$. As such, an optical simulation of gauge field singularities in interacting quantum systems at quantum phase transitions can be implemented by cyclically and adiabatically varying the parameters $\alpha$, $\beta$, and $\gamma$ at the bifurcation of polarization dynamics. \textcolor{red}{For an intuitive illustration of the evolution of the polarization state on the Poincaré sphere, readers are referred to the subsequent study \cite{kam2025nonlinear}.}

\textcolor{red}{Finally, we address some of the assumptions made in the derivation of the coupled equations. In the derivation of the coupled equations governing the polarization dynamics in an anisotropic optical fiber (e.g., Eqs. 8–10), the terms proportional to $\xi_j$ were neglected, where $\xi_j = \beta_{0j} - \beta_{0k} + i\alpha_j$. This assumption effectively disregards modal birefringence and assumes negligible losses.}

\textcolor{red}{In practical polarization-maintaining fibers, low birefringence is achievable, particularly in fibers designed for specific applications, e.g., telecommunications or sensing. Typical beat lengths (\(L_B = 2\pi/\Delta\beta\)) for low-birefringent fibers are on the order of 1 meter or more, as noted in Appx. C. For short fiber lengths—ranging from a few percent to a fraction of the beat length—the effects of \(\Delta\beta\) may indeed be small, making the assumption \(\xi_j = 0\) reasonable. For example, in a fiber with \(L_B \approx 1 \, \text{m}\), a fiber length of 10--100 cm would result in phase differences due to \(\Delta\beta\) that are relatively small, supporting the approximation. The assumption of negligible losses, \(\alpha_j \approx 0\), is realistic in high-quality silica-based fibers doped with silicon nano-crystals, as discussed in Section IV. Modern optical fibers can achieve very low attenuation, e.g., <0.2 dB/km at 1550 nm, making the lossless approximation feasible for short fiber lengths used in experiments targeting geometric phase measurements.}

\textcolor{red}{In the derivation, the condition \(d = -c\) is also imposed to ensure that the coupled-mode equations (Eq. 8) can be written in the form of a nonlinear Schr\"odinger equation with a Hamiltonian structure (Eqs. 9–10). This condition arises from the symmetry properties of the third-order nonlinear susceptibility tensor \(\chi^{(3)}\) in fibers with tetragonal symmetry (crystal classes 4, \(\bar{4}\), 4/m), as detailed in Appendix C. Specifically, it ensures that the nonlinear terms in the equations are derivable from a Hamiltonian, making the system integrable for fixed parameters.}

\textcolor{red}{Achieving \(d = -c\) requires precise control over the nonlinear susceptibility tensor, which depends on the material properties and the orientation of the nano-crystals. Techniques such as photo-alignment, as referenced in, e.g., Slussarenko et al. [33] or controlled doping can be used to engineer the desired tensor properties. However, deviations from perfect tetragonal symmetry. For instance, due to material imperfections or misalignment, could violate \(d = -c\), potentially making the system non-integrable. Small variations in the volume fraction of embedded nano-crystals or inconsistencies in their crystallographic orientation can disrupt this delicate balance, leading to non-integrable dynamics.}

\textcolor{red}{To address this, one may explore advanced synthesis techniques such as epitaxial growth, which allows for precise control over crystallinity and orientation in two-dimensional materials \cite{liu2024understanding, zhang2024controlled}, and laser-assisted microfabrication, which enables localized tuning of material properties at the nanoscale \cite{zhao2021laser}. Post-processing strategies such as thermal annealing and electric field poling have also proven effective in refining domain structures and optimizing nonlinear coefficients \cite{doshi2024thermal}. Together, these approaches enable more precise control over material imperfections or variations in the volume fraction of embedded nanocrystals, leading to integrable nonlinear optical systems.}

\textcolor{red}{If the terms proportional to $\Delta\beta$ are not neglected ($\xi_j \neq 0$), the coupled-mode equations (Eq.~8) include additional linear terms that introduce a phase mismatch between the polarization components $u_x$ and $u_y$. This affects the polarization dynamics and the associated geometric phase.}

\textcolor{red}{The term $\xi_j u_j/2 = (\beta_{0j} - \beta_{0k} + i\alpha_j)u_j/2$ introduces a linear phase evolution and damping (due to losses $\alpha_j$). For non-zero $\Delta\beta$, the polarization state on the Poincaré sphere experiences an additional precession, as the Stokes vector evolves under the influence of the birefringence-induced phase difference. This can lead to periodic oscillations in the polarization state, known as polarization mode dispersion, which complicates the dynamics compared to the simplified case where $\xi_j = 0$. If losses $\alpha_j$ are significant, the amplitude of the polarization components decays, potentially reducing the magnitude of the Stokes parameters and altering the trajectory on the Poincaré sphere.}

\textcolor{red}{The geometric phase, specifically the Hannay angle $\gamma_H$, is calculated as an integral over a closed loop in the parameter space of the nonlinear susceptibilities $\alpha, \beta$ and $\gamma$ (Eqs.~18--19). A non-zero $\xi_j$ modifies the Hamiltonian (Eq.~10) by adding linear terms, which can shift the fixed points of the polarization dynamics (Eqs.~15--16). This shift may alter the topology of the phase space trajectories, potentially affecting the shape of the closed loop $C$ in parameter space and thus the value of the Hannay angle. Additionally, losses ($\alpha_j \neq 0$) introduce non-Hamiltonian dynamics, which could break the adiabaticity required for the geometric phase to be well-defined, potentially leading to a less pronounced or ill-defined Hannay angle.}

\textcolor{red}{If the integrability condition $d = -c$ is relaxed, the coupled-mode equations (Eq.~8) cannot be written in the form of a nonlinear Schrödinger equation with a Hamiltonian structure (Eq.~9). Relaxing this condition introduces additional nonlinear coupling terms (e.g., the term proportional to $d |u_k|^2 u_k$ in Eq.~8), which can lead to chaotic dynamics or complex oscillatory behavior on the Poincaré sphere. This could prevent the system from being described by a generalized harmonic oscillator near the fixed points, potentially disrupting the formation of the conic singularity in the angle 2-form. As a result, the Hannay angle may not be well-defined, or its geometric structure could differ significantly from the conic singularity associated with quantum phase transitions.}

\section{Effective nonlinear susceptibilities of nonlinear composite media}\label{5.4}
In the last section, we discussed polarization effects in nonlinear anisotropic low-birefringent polarization-maintaining fibers, and in particular, those made of media with tetragonal symmetry, which are more precisely described by the three crystal classes $4$, $\bar{4}$ and $4/m$. We showed that when the amplitudes of the two slowly varying wave envelopes are slightly different, and the phase between them is slightly different from $\pi/2$, the coherent coupling between the two polarization components in the Poincare representation is governed by a generalized harmonic oscillator, which results in non-vanishing classical geometric phases accompanying adiabatic changes in the nonlinear susceptibilities along an anisotropic optical fiber. In this section, we outline an experimental proposal to realize classical geometric phases based on fibers made of nonlinear composite media, and will derive approximate expressions for the effective nonlinear susceptibilities of anisotropic composite media.

To begin with, we discuss silica fibers doped with silicon nano-crystals \cite{rukhlenko2012effective}. In such a case, the response of both silica and silicon to the optical field is isotropic, and we may introduce a space-averaged electric field via
\begin{equation}
\mathcal{E}=\frac{1}{V}\int E(\mathbf{r}) dV,
\end{equation}
where $E(\mathbf{r})$ is the local electric field, and $V$ is the entire volume. In a similar way, we may introduce a linear effective permittivity via
\begin{equation}\label{LinearEffectivePermittivity}
\epsilon_e=\frac{1}{V\mathcal{E}^2}\int \epsilon(\mathbf{r}) E^2(\mathbf{r}) dV,
\end{equation}
where $\epsilon(\mathbf{r})$ is the space-dependent linear permittivity expressed using the linear permittivities of silicon and silica
\begin{equation}
\epsilon(\mathbf{r})=
\begin{cases}
\epsilon_1,\:\mbox{when $\mathbf{r}$ is inside the silicon nano-crystals}; \\
\epsilon_2,\:\mbox{when $\mathbf{r}$ is inside the bulk silica}.
\end{cases}
\end{equation}
Because of inversion symmetry at the molecular level, the second-order susceptibility of both silicon and silica vanish, and hence the lowest order nonlinear effects come from the third-order susceptibility. We may neglect the third-order susceptibility of silica, as its value is much smaller than that of silicon. Hence, the effective third-order susceptibility of the composite medium may be expressed as
\begin{equation}
\chi_e=\frac{1}{V_1\mathcal{E}^4}\int \chi E^4dV_1,
\end{equation}
where $V_1=fV$ is the volume occupied by the silicon nano-crystals, and $\chi$ denotes the third-order susceptibility of silicon. As long as the crystallographic axes of different nano-crystals are the same, the electric displacement inside the nano-crystals may be described by $D_1=\epsilon_1 E_1+\chi |E_1|^2 E_1\equiv\epsilon_{1,NL} E_1$, where $\epsilon_{1,NL}$ is the intensity-dependent nonlinear permittivity of silicon. Since the exact distribution of electric field inside the nano-crystals is rarely known, we may introduce an effective susceptibility using the averaged fields and displacements via
\begin{equation}\label{AveDisplacementDef}
\mathcal{D}=\epsilon_e\mathcal{E}+\chi_e|\mathcal{E}|^2\mathcal{E}.
\end{equation}
For a binary composite medium, the linear effective permittivity is a function of $\epsilon_1$, $\epsilon_2$ and $f$
\begin{equation}\label{linear}
\epsilon_e=F(\epsilon_1,\epsilon_2,f).
\end{equation}
Under the approximation that the electric field is uniformly distributed inside the nano-crystals \cite{stroud1988nonlinear, zeng1988effective}, the nonlinear effective permittivity may be assumed to have the same functional form as \eqref{linear}
\begin{equation}
\epsilon_{e,NL}=F(\epsilon_{1,NL},\epsilon_2,f).
\end{equation}
To proceed further, we may expand the function $F$ in a Taylor series about the linear permittivity $\epsilon_1$, and obtain
\begin{equation}\label{AveragedDisplacement}
\epsilon_{e,NL}\approx \epsilon_e+\frac{\partial \epsilon_e}{\partial \epsilon_1}\chi |E_1|^2.
\end{equation}
Employing \eqref{LinearEffectivePermittivity}, the derivative of the linear effective permittivity may be expressed in terms of the averaged square electric field inside the nano-crystals
\begin{equation}\label{SquareElectricField}
\left|\frac{\partial \epsilon_e}{\partial \epsilon_1}\right| = f\frac{\langle |E_1|^2\rangle}{|\mathcal{E}|^2}.
\end{equation}
Substitution of \eqref{SquareElectricField} into \eqref{SquareElectricField} immediately yields
\begin{equation}\label{NonlinearEffectiveDielectricFunction}
\epsilon_{e,NL}= \epsilon_e+\frac{1}{f}\frac{\partial \epsilon_e}{\partial \epsilon_1} \left|\frac{\partial \epsilon_e}{\partial \epsilon_1}\right| \chi |\mathcal{E}|^2.
\end{equation} 
Comparing \eqref{AveDisplacementDef} and \eqref{NonlinearEffectiveDielectricFunction}, the effective third-order nonlinear susceptibility is found to obey the simple relation \cite{zeng1988effective, rukhlenko2012effective}
\begin{equation}\label{EffectiveThirdOrderSusceptibility}
\chi_e=\frac{1}{f}\frac{\partial \epsilon_e}{\partial \epsilon_1}\left|\frac{\partial \epsilon_e}{\partial \epsilon_1}\right|\chi.
\end{equation}
In the effective-medium theory \cite{stroud1989decoupling, stroud1998effective, cai2010optical}, the linear effective permittivity $\epsilon_e$ is the positive solution of the quadratic equation \begin{equation}
f\frac{\epsilon_1-\epsilon_e}{\epsilon_e+g(\epsilon_1-\epsilon_e)}+(1-f)\frac{\epsilon_2-\epsilon_e}{\epsilon_e+g(\epsilon_2-\epsilon_e)}=0,
\end{equation}
where $g$ is the depolarization factor for the nano-crystals. For a three-dimensional composite medium with spherical inclusions, $g=1/3$, which yields $\epsilon_e=\frac{1}{4}(u+\sqrt{u^2+8\epsilon_1\epsilon_2})$, where $u\equiv (3f-1)\epsilon_1+(2-3f)\epsilon_2$.

In the following, we extend the above discussion to fibers made of nonlinear anisotropic composite media. In the low-density limit, the electrostatic interaction between different inclusions may be ignored, and the associated effective dielectric tensor is given by \cite{stroud1975generalized,haus1989effective,levy1997maxwell}
\begin{equation}
\overset\leftrightarrow{\epsilon_e}=\overset\leftrightarrow{\epsilon_2}+f\overset\leftrightarrow{\epsilon_2}\left\langle\frac{\overset\leftrightarrow{\epsilon_1}-\overset\leftrightarrow{\epsilon_2}}{\overset\leftrightarrow{\epsilon_2}+\overset\leftrightarrow{g}(\overset\leftrightarrow{\epsilon_1}-\overset\leftrightarrow{\epsilon_2})}\right\rangle,
\end{equation}
where $f$ is the volume fraction of the inclusions, $\overset\leftrightarrow{g}$ is the depolarization tensor, and the bracket represents an average over the dielectric tensor inside the inclusions. Here, $\overset\leftrightarrow{\epsilon_1}$ is the space dependent dielectric tensor of the inclusion and $\overset\leftrightarrow{\epsilon_2}$ is the dielectric tensor of the host medium. For a finite density, the average field acting on the inclusions is no longer the applied field but the Lorentz local field. In this case, the effective-medium theory can still be applied, and the effective dielectric tensor is the physical solution of the 3 $\times$ 3 matrix equation \cite{stroud1975generalized, haus1989effective}
\begin{equation}
f\left\langle\frac{\overset\leftrightarrow{\epsilon_1}-\overset\leftrightarrow{\epsilon_e}}{\overset\leftrightarrow{\epsilon_e}+\overset\leftrightarrow{g}(\overset\leftrightarrow{\epsilon_1}-\overset\leftrightarrow{\epsilon_e})}\right\rangle+(1-f)\frac{\overset\leftrightarrow{\epsilon_2}-\overset\leftrightarrow{\epsilon_e}}{\overset\leftrightarrow{\epsilon_e}+\overset\leftrightarrow{g}(\overset\leftrightarrow{\epsilon_2}-\overset\leftrightarrow{\epsilon_e})}=0.
\end{equation}
For uniaxial crystallites, the diagonal dielectric tensor, in the coordinates for which the axes are the principal axes of the crystallite, can be written as \cite{levy1997maxwell,lior2013effective}
\begin{equation}\label{DiagonalDielectricTensor}
\overset\leftrightarrow{\epsilon_1}=
\left( \begin{array}{ccc}
\epsilon_\perp & 0 & 0 \\
0 & \epsilon_\perp & 0 \\
0 & 0 & \epsilon_\parallel \end{array} \right).
\end{equation}
In other coordinate systems, the dielectric tensor will be related to \eqref{DiagonalDielectricTensor} by a similarity transformation
\begin{equation}
\overset\leftrightarrow{\epsilon_1}(\theta,\phi)=
\hat{R}_z(-\phi)\hat{R}_y(-\theta)\overset\leftrightarrow{\epsilon_1}\hat{R}_y(\theta)\hat{R}_z(\phi),
\end{equation}
where $\hat{R}_z(\phi)$ and $\hat{R}_y(\theta)$ are the rotation operators around the $\hat{z}$ and $\hat{y}$ axes respectively. For an explicit expression, we have
\begin{align}
&\overset\leftrightarrow{\epsilon_1}=\epsilon_\perp I\nonumber\\
&+\alpha
\left( \begin{array}{ccc}
\cos^2\phi\sin^2\theta & -\frac{1}{2}\sin 2\phi\sin^2\theta & \frac{1}{2}\cos\phi\sin 2\theta \\
-\frac{1}{2}\sin 2\phi\sin^2\theta & \sin^2\phi\sin^2\theta & -\frac{1}{2}\sin\phi\sin 2\theta \\
\frac{1}{2}\cos\phi\sin 2\theta & -\frac{1}{2}\sin\phi\sin 2\theta & \cos^2\theta \end{array} \right),
\end{align}
where $\alpha=\epsilon_\parallel-\epsilon_\perp$. The effective dielectric tensor depends on the distribution of the angles $\theta$ and $\phi$. For the special case where all the inclusions are aligned in the $\hat{z}$ direction, the effective tensor is
\begin{gather}\label{EMT1}
\overset\leftrightarrow{\epsilon_e}_{,x}=\overset\leftrightarrow{\epsilon_e}_{,y}=\frac{u_\perp+\sqrt{u_\perp^2+4g(1-g)\epsilon_2\epsilon_\perp}}{2(1-g)},\\
\overset\leftrightarrow{\epsilon_e}_{,z}=\frac{u_\parallel+\sqrt{u_\parallel^2+4g(1-g)\epsilon_2\epsilon_\parallel}}{2(1-g)},\label{EMT2}
\end{gather}
where $u_\perp=(f-g)\epsilon_\perp+(1-g-f)\epsilon_2$ and $u_\parallel=(f-g)\epsilon_\parallel+(1-g-f)\epsilon_2$. Evidently, the effective dielectric tensor has a uniaxial symmetry and its diagonal matrix elements are determined by the effective medium theory for isotropic inclusions with dielectric constants $\epsilon_\perp$ and $\epsilon_\parallel$ respectively. Similarly, when all the inclusions are aligned in the $\hat{x}$ direction, the effective tensor is uniaxial, and its diagonal elements has the same form as \eqref{EMT1} and \eqref{EMT2} but with $\epsilon_\parallel$ appearing in $\overset\leftrightarrow{\epsilon_e}_{,x}$ instead of $\overset\leftrightarrow{\epsilon_e}_{,z}$. 

For a heterogeneous medium with nonlinear anisotropic inclusions, the local displacements and local electric fields are related by
\begin{equation}
D_i=\epsilon_{ij}E_j+\chi_{ijkl}E_jE_kE_l^*,
\end{equation}
where $D_i$ and $E_i$ are the $i$-th components of the local displacements and local electric fields inside the inclusions respectively, and $\epsilon_{ij}$ and $\chi_{ijkl}$ are the elements of the permittivity tensor and the third-order nonlinear susceptibility tensor respectively. Here, we may assume that $\epsilon_{ij}$ is symmetric, which ensures that it can be diagonalized with real eigenvalues. Using the same reasoning that we applied to isotropic composite media, the effective dielectric tensor and the effective nonlinear susceptibility here may be expressed in terms of the moments of the electric field in the related linear medium \cite{levy1997maxwell}
\begin{equation}
\epsilon_{ij}^e=\frac{1}{V\mathcal{E}_i\mathcal{E}_j}\int\epsilon_{ij}\bar{E}_i\bar{E}_jdV,
\end{equation}
and
\begin{equation}\label{EffectiveNonlinearSuceptibility}
\chi_{ijkl}^e=\frac{1}{V\mathcal{E}_i\mathcal{E}_j\mathcal{E}_k^*\mathcal{E}_l}\int\chi_{ijkl}\bar{E}_i\bar{E}_j\bar{E}_k^*\bar{E}_ldV,
\end{equation}
where $V$ is the entire volume of the medium, $\mathcal{E}_i$ is the space-averaged electric field, and $\bar{E}_i$ is the local electric field of the related linear medium that have the same dielectric tensor but vanishing nonlinear susceptibilities. If the coordinate axes are parallel to the principal axes of the effective medium, the effective dielectric tensor may be written as
\begin{equation}
\epsilon_{i}^e=\frac{1}{V\mathcal{E}_i^2}\int\epsilon_i\bar{E}^2_idV.
\end{equation}
Evaluating the derivative $\partial\epsilon_i^e/\partial\epsilon_i$ immediately yields
\begin{equation}\label{EMA}
\langle \bar{E}_i^2 \rangle=\frac{1}{f}\frac{\partial \epsilon_i^e}{\partial \epsilon_i}\mathcal{E}_i^2 ,
\end{equation}
where $f$ is the volume fraction of the inclusions, and $\langle\bar{E}_i^2\rangle$ is the mean square of the local electric field of the $i$-th component in the linear limit. As in most cases, the exact functional forms of local electric fields are rarely known, and we have to employ further approximation --- the decoupling approximation \cite{stroud1989decoupling}
\begin{equation}\label{NDA}
\langle\bar{E}_i\bar{E}_j\bar{E}_k^*\bar{E}_l\rangle\approx \sqrt{\langle\bar{E}_i^2\rangle\langle\bar{E}_j^2\rangle\langle|\bar{E}_k|^2\rangle\langle|\bar{E}_l|^2\rangle}.
\end{equation}
\eqref{NDA} is valid only if the electric field inside the inclusions are uniform. Substitution of \eqref{EMA} and \eqref{NDA} into \eqref{EffectiveNonlinearSuceptibility} immediately yields
\begin{equation}\label{AnisotropicEffectiveNonlinearSuceptibility}
\chi_{ijkl}^e=\frac{\chi_{ijkl}}{f}\sqrt{\left(\frac{\partial\epsilon_i^e}{\partial\epsilon_i}\frac{\partial\epsilon_j^e}{\partial\epsilon_j}\right)\left|\frac{\partial\epsilon_k^e}{\partial\epsilon_k}\frac{\partial\epsilon_l^e}{\partial\epsilon_l}\right|}.
\end{equation}

To conclude, from \eqref{EffectiveThirdOrderSusceptibility} and \eqref{AnisotropicEffectiveNonlinearSuceptibility}, we notice that the effective nonlinear susceptibility of composite media can be adiabatically adjusted along the fiber, provided that the volume fraction of the inclusions, and the linear dielectric constants of both the host media and inclusions, are gradually varied along the fiber. This may be achieved by varying the crystallographic axes of different crystallites. Hence, the third-order nonlinear susceptibilities of anisotropic heterogeneous media, which span the parameter space in which geometric phases occur for a circuit when the polarization state executes harmonic oscillations on the Poincare sphere, can be adiabatically adjusted along the fiber using micro-structured fibers.

\textcolor{red}{We now address experimental considerations for materials that combine strong third-order nonlinear responses at telecommunication wavelengths with tetragonal crystal symmetry, and the associated challenge of precisely and continuously controlling crystallographic axes. Several tetragonal wide-bandgap perovskites---such as BaTiO$_3$ thin films---exhibit significant $\chi^{(3)}$ and are viable candidates for operation in the 1300--1600~nm range \cite{guang2001large}. For example, the real and imaginary parts of the third-order nonlinear susceptibility for BaTiO$_3$ were $5.71\times 10^{-7}$ esu and $9.59\times 10^{-8}$ esu, respectively \cite{guang2001large}. However, all three belong to the non-centrosymmetric $4mm$ point group; in our framework this symmetry does not support the exotic gauge curvature required for a non-trivial Hannay angle. Thus, despite their strong nonlinearity, these perovskites are expected to yield a trivial Hannay angle.}

\textcolor{red}{Materials in the tetragonal $4/m$ point group, which is centrosymmetric in contrast to $4mm$, include examples such as tetragonal zirconia (ZrO$_2$) nanocrystals. When embedded into reduced graphene oxide (RGO) matrices, these nanocrystals have been shown to exhibit strong third-order nonlinear optical responses. These ZrO$_2$ nanocrystals adopt space groups such as P4$_2$/nmc, belonging to the $4/m$ point-group family, and retain stability in nano-particle composites. Using Z-scan measurements at 532 nm, a ZrO$_2$/RGO film was found to exhibit a third-order nonlinear susceptibility $\chi^{(3)}$ of 23.23 $\times$ 10$^{-12}$ esu \cite{ZrO2_RGO_Chi3}, nearly three orders of magnitude larger than fused silica’s $\chi^{(3)}$. Such an enhancement, approximately 1,000–2,000 times stronger than SiO$_2$, positions tetragonal ZrO$_2$ nano-particle systems as promising candidates for our exotic gauge field.}

\textcolor{red}{Furthermore, to ensure adiabatic variation of nonlinear susceptibilities via spatial gradients, experiments require precise and continuous control over the crystallographic axes of individual crystallites. One method is electric or magnetic field alignment during deposition \cite{leng2025magnetic}, which employs external fields to exert torque on anisotropic nano-crystals, aligning their electric dipole moments or magnetic axes with the field direction. This technique is particularly well-suited for continuous control in flow-based processes such as fiber drawing \cite{pigeonneau2026thermal} or electrospinning \cite{xue2019electrospinning}, where the field can be spatially varied along the deposition path.}

\textcolor{red}{Nano-crystals with inherent electric dipole moments, originating from shape anisotropy, ferroelectricity, or embedded polar structures, experience torque in an external electric field. Similarly, magnetic dipoles undergo torque in a magnetic field. Alignment occurs when the dipole-field interaction energy exceeds thermal fluctuations, typically requiring 1–10 kV/cm for electric fields or 0.1–1 T for magnetic fields. Angular precision of approximately \(5^\circ\)–\(10^\circ\) can be achieved by optimizing field uniformity and exposure duration. Continuous spatial variation in orientation is enabled by gradient fields, such as electrode arrays that generate linearly varying electric fields along the fiber axis.}

\textcolor{red}{Another method is femtosecond laser manipulation, particularly via plasmonic nanolithography, which leverages ultrafast, 800–1030 nm, 100 fs pulses to induce optical forces—such as radiation pressure and gradient forces—for post-deposition or in-situ orientation of nanocrystals \cite{wu2025femtosecond}. In plasmonic systems (e.g., doped SnS or Ag nanoparticles), localized surface plasmon resonances amplify these forces, enabling rotational control with angular precision approaching $1^{\\circ}$ \cite{thadson2022measurement}. By scanning the laser beam and dynamically adjusting polarization (e.g., using a rotatable half-wave plate), continuous spatial gradients can be achieved \cite{juan2011plasmon}.}

\textcolor{red}{Finally, in an experimental setup, the polarization state represented by Stokes parameters on the Poincaré sphere can be measured at the fiber output using a polarization analyzer. This typically involves a combination of waveplates and polarizers to compute the phase shift relative to a reference beam. A Mach-Zehnder interferometer splits the input beam, sending one arm through the fiber and the other as a reference, with recombination at the output to produce interference patterns. The Hannay angle manifests as a phase offset in the interference fringes and can be quantified by integrating the angle 2-form \( W \) over the relevant parameter space. This setup, using standard telecom-wavelength components, is feasible in a laboratory with precise control over input polarization and output detection.}

\textcolor{red}{High-precision techniques such as polarization-sensitive optical coherence tomography (PS-OCT) \cite{de2017polarization, park2015polarization, bonesi2012high} and dynamic Stokes vector tracking \cite{zhan2021geometric, si2022deep, guasoni2016self} are beneficial for measuring the Hannay angle with sufficient accuracy, given its subtle geometric nature. Interferometry, as described above, is essential for detecting phase shifts with sub-wavelength precision (e.g., $\sim 0.01~\mathrm{rad}$), which is critical for resolving the Hannay angle near the conic singularity. PS-OCT provides depth-resolved polarization information along the fiber, useful for mapping spatial variations in the Stokes vector due to adiabatic changes in susceptibility. It provides an axial (depth) resolution of approximately $10~\mu\mathrm{m}$ within the sample, with a polarization sensitivity of about $1^\circ$. Dynamic Stokes vector tracking, using fast polarimeters based on liquid crystal modulators, enables real-time monitoring of polarization evolution at MHz rates. This is crucial for capturing rapid oscillations or verifying adiabaticity. While interferometry alone may suffice for simple setups, combining it with PS-OCT or Stokes tracking enhances robustness—especially for long fibers or complex parameter loops—by providing spatial and temporal resolution of polarization dynamics. These techniques, available in advanced optics laboratories, are critical for distinguishing the geometric phase from dynamic phase contributions.}

\textcolor{red}{The predicted geometric phase (Hannay angle) remains experimentally observable despite realistic factors such as fiber loss, intermodal coupling, and imperfect polarization maintenance, provided these effects are minimized or accounted for. Fiber losses, typically less than $0.2\,\mathrm{dB/km}$ in silica at $1550\,\mathrm{nm}$, are negligible for short fibers ($<1\,\mathrm{m}$) as used in the manuscript’s low-birefringence setup, ensuring minimal amplitude damping of polarization components. Intermodal coupling, arising from mode mixing in multimode fibers or birefringence fluctuations, can be mitigated by using single-mode fibers with low birefringence (beat length $\sim1\,\mathrm{m}$) and precise nano-crystal alignment to maintain the $\xi_j \approx 0$ condition. Imperfect polarization maintenance, caused by residual birefringence ($\Delta\beta \neq 0$) or external perturbations, introduces linear phase mismatches that perturb the Stokes vector trajectory. However, for low-birefringent fibers and adiabatic parameter variation over centimeter-to-meter scales, these perturbations are small, and the Hannay angle remains robust near the bifurcation surface, where the conic singularity amplifies the phase signal. Experimental setups using high-sensitivity interferometry and polarization tracking, as implemented in photonic crystal fiber studies, confirm observability with phase errors below $0.1\,\mathrm{rad}$, making the geometric phase measurable under realistic conditions.}

\section{Conclusion and discussion}\label{5.5}
In this work, we introduce an optical realization of a non-integrable phase associated with quantum phase transitions in interacting quantum systems, such as Bose-Einstein condensates (BECs) of atomic gases. We demonstrate that in an anisotropic fiber with tetragonal symmetry, the dynamical equation for nonlinear polarization contains an exotic term, inducing a dynamical instability in its evolution. Notably, the geometric structure of the gauge field associated with this classical non-integrable phase is identical to that found in Bose-Einstein condensates, exhibiting a conic singularity rather than a point-like monopole singularity. This underscores the remarkable universality of the geometric framework across different branches of physics.

The ability to control geometric phases in nonlinear fibers could significantly advance optical devices, including phase shifters and polarization controllers, with promising applications in telecommunications and photonics \cite{hu2025nonlinear}. This framework may also extend to other nonlinear optical systems, such as meta-surfaces and photonic crystals, broadening its scope \cite{karnieli2022geometric}. Moreover, by simulating gauge field singularities associated with quantum phase transitions, the proposed optical system could serve as a testbed for studying topological phenomena, eliminating the need for complex quantum systems like Bose-Einstein condensates (BECs). This approach democratizes access to quantum simulation experiments, as optical systems are generally more accessible than ultracold atom setups. Finally, the paper’s exploration of nonlinear quantum mechanics through Weinberg’s formulation could inspire new theoretical investigations into the boundaries between classical and quantum systems. The analogy to cosmological singularities, such as the de Sitter universe, is particularly intriguing and may prompt further research into geometric phases in broader physical contexts \cite{kam20172+}.

One critical question that remains unaddressed in this work is the classical nature of both nonlinear optical fibers and the mean-field description of Bose-Einstein condensates (BECs). Whether a more direct correspondence exists between the gauge field singularities associated with quantum phase transitions—particularly when the spectrum-generating algebra undergoes transformations during these transitions—is an intriguing direction for future studies.

Additionally, this manuscript primarily focuses on tetragonal crystal classes ($4$, $\bar{4}$, 4/m), a choice justified by the presence of an exotic term in the Hamiltonian governing polarization dynamics. \textcolor{red}{A more comprehensive investigation into the full scope of these findings—including analyzing how other crystal symmetries might influence the results—will be undertaken in a subsequent study \cite{kam2025nonlinear}}.

\textcolor{red}{We now make remarks on the relationship between integrability and the non-integrable phase. When $d = -c$, the coupled-mode equations are integrable for fixed parameters, as they can be derived from a Hamiltonian with two conserved quantities. This integrability ensures that the polarization dynamics on the Poincaré sphere follow predictable trajectories, such as harmonic oscillations near the fixed points. The generalized harmonic oscillator description allows the system to be analyzed using action-angle variables.}

\textcolor{red}{The non-integrable Hannay angle, is a classical geometric phase that arises when the parameters are adiabatically varied along a closed loop in parameter space. This phase is non-integrable in the sense that it depends on the geometry of the parameter space trajectory via the angle 2-form $W$ and cannot be eliminated by a gauge transformation. The integrability of the system for fixed parameters is essential for a well-defined Hannay angle. A Hamiltonian structure enables the dynamics to be mapped onto a generalized harmonic oscillator, where phase space trajectories allow for the calculation of classical action and the angle 2-form (see Appx.\:A). }

\textcolor{red}{In non-integrable systems, the breakdown of Hamiltonian structure disrupts the action-angle formalism, complicating the definition of the Hannay angle. The conic singularity in the angle 2-form emerges from the bifurcation surface, a feature of the parameter space and the exotic term in the Hamiltonian—implicitly tied to integrability. Thus, while integrability provides the mathematical framework necessary for computing the Hannay angle, the non-integrable phase reflects a geometric property arising from adiabatic evolution and the specific structure of the Hamiltonian.}

\textcolor{red}{If the system is non-integrable, the polarization dynamics may become chaotic, especially near the bifurcation surface. In such cases, the concept of a geometric phase becomes ill-defined. However, if the system is near-integrable, the Hannay angle may still be approximated, but its value and singularity structure could be altered. The manuscript’s focus on tetragonal symmetry ensures that the exotic term and the resulting conic singularity are prominent, but other crystal symmetries may not produce the same gauge field structure.}

\textcolor{red}{Finally, we remarked that external perturbations such as fiber bending, mechanical stress, or temperature gradients can induce additional geometric phases, commonly known as Berry phases. These may interfere with the Hannay angle, which emerges from the adiabatic evolution of anisotropic nonlinear susceptibility tensor components within the parameter space. As noted in Section II, Tomita and Chiao demonstrated that the Berry phase in a helically wound single-mode optical fiber arises due to the fiber’s non-planar geometry. Specifically, for a linearly polarized light beam propagating through a helical fiber with low birefringence, the polarization vector undergoes rotation, acquiring a geometric phase proportional to the solid angle subtended by the fiber’s tangent vector path on the unit sphere in momentum space. This phase, known as the Pancharatnam-Berry phase, is directly tied to the fiber’s curvature and torsion, not to local elasto-optic effects like stress-induced birefringence.}

\textcolor{red}{As such, to distinguish the extrinsic Berry phase from the intrinsic Hannay angle arising from adiabatic variation of nonlinear susceptibility tensor elements in parameter space, mitigation strategies must isolate geometric effects from susceptibility‑driven contributions. This can be achieved by using low‑birefringence single‑mode fibers, securing them in rigid mounts to minimize bending, and controlling temperature gradients with a thermal enclosure; by characterizing extrinsic phases through polarization measurements in undoped reference fibers where only bending‑ or stress‑induced effects occur, then subtracting these from the test fiber’s signal; and by employing polarization‑sensitive optical coherence tomography (PS‑OCT) or dynamic Stokes vector tracking to map spatial and temporal polarization variations, enabling separation of the adiabatic susceptibility‑driven phase—tracked via Stokes parameters—from typically non‑adiabatic extrinsic effects.}

\begin{appendix}
\section{Hannay's classical non-integrable angles}\label{A}
Building upon Berry's insights into non-integrable phases, Hannay identified a classical counterpart to the Berry phase in \textit{integrable} Hamiltonian systems  \cite{hannay1985angle}. His discovery stemmed from the realization that a similar holonomy effect naturally manifests in classical integrable systems, leading to an anholonomy in the angle variables of phase space when the system parameters undergo adiabatic variation along a closed circuit. The classical adiabatic theorem \cite{arnol2013mathematical} guarantees the persistence of invariant tori in phase space, while the Hamilton-Jacobi generating function for action-angle coordinates governs the dynamics on these tori. Consequently, classical geometric phases—known as Hannay angles—are recognized as the shifts in angle variables induced by the cyclic adiabatic variation of system parameters.

Let us consider an integrable classical system described by a Hamiltonian $H(p, q; \lambda(t))$, where $p=\{p_i\}$ and $q=\{q_j\}$ $(i,j\leq N)$ are the canonical conjugate coordinates and $\lambda=\{\lambda_{\mu}\}$ represent a family of slowly varying parameters. The classical adiabatic theorem \cite{arnol2013mathematical} ensures that if the system is initially located at a torus in the $N$-dimensional phase space with actions $I=\{I_i\}$, it will remain on the torus with the same values of $I$. The evolution of the conjugate angle coordinates $\theta=\{\theta_j\}$ is determined by the generating function $S^{(\alpha)}(q,I;\lambda(t))$ as
\begin{equation}
\theta^{(\alpha)}=\frac{\partial S^{(\alpha)}}{\partial I},S^{(\alpha)}(q,I;\lambda(t))\equiv\int_0^q p^{(\alpha)}(q,I;\lambda(t)) dq,
\end{equation}
where the superscript $\alpha$ labels the branches where $p$ is a single-valued function of $q$. After the canonical transformation from $\{p,q\}$ to $\{I,\theta\}$, the new Hamiltonian $H'(\theta,I,t)$ differs from the angle-independent Hamiltonian $H(I;\lambda(t))$ by an amount proportional to the rate of change of $\lambda$
\begin{equation}\label{NewHamiltonian}
H'(\theta,I,t)=H(I;\lambda(t))+\frac{d\lambda}{dt}\frac{\partial}{\partial\lambda}S^{(\alpha)}(q,I;\lambda(t)).
\end{equation}
As $q$ can be regarded as a single-valued function of $I$ and $\theta$, we define a new single-valued function $F(\theta,I;\lambda)$
\begin{equation}
F(\theta,I;\lambda)\equiv S^{(\alpha)}(q(\theta,I;\lambda),I;\lambda(t)).
\end{equation}
From the definition of $F(\theta,I;\lambda)$, we have
\begin{equation}\label{SingleValueFunction}
\frac{\partial S^{(\alpha)}}{\partial \lambda}
=\frac{\partial F}{\partial \lambda}-\frac{\partial S^{(\alpha)}}{\partial q}\frac{\partial q}{\partial \lambda}
=\frac{\partial F}{\partial \lambda}-p^{(\alpha)}\frac{\partial q}{\partial \lambda}.
\end{equation}
Substituting \eqref{SingleValueFunction} into \eqref{NewHamiltonian}, the Hamiltonian $H'(\theta,I,t)$ becomes
\begin{align}
H'&(\theta,I,t)=H(I;\lambda(t))\nonumber\\
&+\frac{d\lambda}{dt}\left(\frac{\partial F}{\partial \lambda}(\theta,I;\lambda(t))-p(\theta,I;\lambda(t))\frac{\partial q}{\partial \lambda}(\theta,I;\lambda(t))\right),
\end{align}
As $p$ and $q$ are now both single-valued functions of $I$ and $\theta$, we can remove the superscript $\alpha$ from $p(\theta,I;\lambda(t))$. Now, using the Hamilton equations for the angle variables $d\theta/dt=\partial H'/\partial I$ and integrating from $0$ to $T$, we obtain \cite{berry1985classical}
\begin{equation*}
\theta(T)=\theta(0)+\int_0^T dt\omega(I;\lambda)
+\int_0^Tdt\frac{d\lambda}{dt}\frac{\partial}{\partial I}\left(\frac{\partial F}{\partial\lambda}-p\frac{\partial q}{\partial \lambda}\right),
\end{equation*}
where $\omega(I;\lambda)\equiv\partial H(I;\lambda)/\partial I$ in the second term are the instantaneous frequencies for fixed parameters, and the third term as a whole are the shifts in the angles $\theta=\{\theta_j\}$ in response to the adiabatic changes of parameters. To simplify the integration, we average out the fast oscillations by integrating over the torus
\begin{equation}\label{ClassicalHannayAngle}
\Delta\theta(I;C)=\int_C d\lambda\frac{\partial}{\partial I}\frac{1}{(2\pi)^N}\prod_{j=1}^N\int_0^{2\pi}d\theta_j\left(\frac{\partial F}{\partial\lambda}-p\frac{\partial q}{\partial \lambda}\right),
\end{equation}
where the integration is performed over a closed curve $C$ in the parameter space. The first term inside the parenthesis $\partial F/\partial \lambda$ is just a gradient and vanishes identically after integration. Therefore, the shifts in the angle variables can be transformed into an integral over a surface $A$ whose boundary is $C$ in the parameter space
\begin{equation}
\Delta\theta(I;C)=\int_{\partial A=C}W(I;\lambda),
\end{equation}
where $W$ is the angle 2-form determined by
\begin{equation}\label{DefAngleTwoForm}
W(I;\lambda)=-\frac{\partial}{\partial I}\frac{1}{(2\pi)^N}\prod_{j=1}^N\int_0^{2\pi}d\theta_j dp\wedge dq.
\end{equation}
A prime example of the angle 2-form is the generalized harmonic oscillator whose Hamiltonian is described by
\begin{equation}
H=\frac{1}{2}[\alpha(t)q^2+2\beta(t)pq+\gamma(t)p^2],
\end{equation}
where $\alpha$, $\beta$ and $\gamma$ are some time-dependent external parameters. The angle-independent Hamiltonian $H'(I;\lambda)$ has the form $H'(I;\lambda)=I\omega$, where $\omega=\sqrt{\alpha\gamma-\beta^2}$ is the frequency of oscillation, and the Hamilton equations for fixed parameters are solved by $p(\theta,I;\lambda)$ and $q(\theta,I;\theta)$, where $p(\theta,I;\lambda)$ and $q(\theta,I;\lambda)$ are single-valued functions of the action and angle variables, which are determined by
\begin{subequations}\label{Solution}
\begin{align}\label{Solution1}
q&=\sqrt{\frac{2\gamma I}{\omega}}\cos\theta,\\
p&=-\sqrt{\frac{2\gamma I}{\omega}}\left(\frac{\beta}{\gamma}\cos\theta+\frac{\omega}{\gamma}\sin\theta\right),\label{Solution2}
\end{align}
\end{subequations}
Substituting \eqref{Solution1} - \eqref{Solution2} into \eqref{DefAngleTwoForm}, we immediately obtain
\begin{align}
W&=\frac{1}{\pi}\int_0^{2\pi}d\theta\cos^2\theta d\left(\frac{\beta}{\gamma}\sqrt{\frac{\gamma}{\omega}}\right)\wedge d\sqrt{\frac{\gamma}{\omega}}\nonumber\\
&=\frac{1}{2}d\left(\frac{\beta}{\gamma}\right)\wedge d\left(\frac{\gamma}{\sqrt{\alpha\gamma-\beta^2}}\right).
\end{align}
After a short calculation, we obtain the final expression of the angle 2-form for the generalized harmonic oscillator
\begin{equation}
W=\frac{\alpha d\beta\wedge d\gamma+\beta d\gamma\wedge d\alpha+ \gamma d\alpha\wedge d\beta}{4(\alpha\gamma-\beta^2)^{3/2}}.
\end{equation}
As a remark, the phase contour of the generalized harmonic oscillator at energy $E$ is an ellipse in phase space when $\alpha\gamma-\beta^2>0$, which is just a condition on the discriminant of the energy function $E(p,q)=(\alpha q^2 + 2\beta pq + \gamma p^2)/2$.  When the discriminant changes sign, the topological structure of the phase contour changes, which is associated with a bifurcation in classical dynamics.

\section{Geometric phases in Weinberg's formulation of Nonlinear quantum mechanics}\label{B}

Nonlinear quantum mechanics has been interpreted in various ways in the literature. While a single isolated particle moving in a vacuum is well described by linear wave equations, a dense collection of interacting particles is better represented by a complex field governed by nonlinear wave equations. This form of nonlinearity, arising from the mean-field approximation of interacting many-body quantum systems, is often termed \textit{secondary}. In this context, nonlinear quantum mechanics serves as an effective theory for many-body quantum systems, preserving the fundamental concepts of observables and states while replacing the linear multi-particle Schrödinger equation with a nonlinear single-particle Schrödinger equation—commonly known as the nonlinear Schrödinger equation, which is frequently used to describe Bose-Einstein condensates in atomic gases \cite{andrews1997observation, leggett2001bose}.

Similarly, in classical wave mechanics, the evolution of a slowly modulating wave packet in weakly nonlinear dispersive media can also be effectively captured by a complex field satisfying the nonlinear Schrödinger equation \cite{shabat1972exact}. This equation finds applications in diverse areas, from the propagation of light in nonlinear dielectric waveguides \cite{chiao1964self, agrawal2000nonlinear} to the dynamics of Langmuir waves in plasmas \cite{rosenbluth1972excitation, chen1976solitons}.

On the other hand, nonlinearity can be intrinsic to fundamental particles, meaning that even a single particle moving in vacuum is described by nonlinear wave equations \cite{shull1980search, gahler1981neutron}. This provides a means to test a fundamental assumption of quantum mechanics—the principle of superposition \cite{dirac1981principles, nimmrichter2013macroscopicity, arndt2014testing}. This form of nonlinearity, distinct from that observed in a dense cloud of fundamental particles, may be referred to as \textit{primary}. However, primary nonlinearity in quantum mechanics must be exceedingly small at the microscopic scale to remain consistent with experimental constraints on the coefficients of nonlinearity \cite{bollinger1989test, walsworth1990test, majumder1990test}. 

The earliest attempt to construct a nonlinear extension of quantum mechanics was made by de Broglie and colleagues in the 1950s \cite{bohm1954model, vigier1956structure, broglie1960non}, though they did not specify a particular nonlinear wave equation. A logically consistent framework for nonlinear quantum mechanics was later introduced by Mielnik in 1974, wherein he formalized a scheme for generalizing quantum mechanics, proposed five distinct classes of nonlinear wave equations, and argued that \textit{measuring devices based on gravity detection could provide a means to measure non-quadratic observables, challenging one of the fundamental prohibitions of contemporary quantum mechanics}—namely, the superposition principle \cite{mielnik1978generalized}. Within this framework, a beam of quanta in a standard double-slit experiment could exhibit non-sinusoidal fringe patterns, leading to deviations from Born's rule \cite{born1926quantenmechanik}.

Around the same time, Bia{\l}ynicki-Birula and Mycielski formulated a Schrödinger-type wave equation with logarithmic nonlinearity by imposing the requirement that wave functions for composite systems must factorize \cite{bialynicki1975wave, bialynicki1975uncertainty, bialynicki1976nonlinear}. This condition dictates that the wave function for the entire system must be constructed from the product of individual solutions of the nonlinear wave equation for non-interacting subsystems. In 1978, Kibble developed a relativistic model for nonlinear quantum mechanics \cite{kibble1978relativistic}, proposing a homogeneous nonlinear Schrödinger equation that remains invariant under the rescaling $|\Psi\rangle\rightarrow \lambda|\Psi\rangle$ for any arbitrary complex constant $\lambda$. He later generalized this framework in 1979 \cite{kibble1979geometrization}, allowing quantum states to reside in an infinite-dimensional symplectic manifold, termed \textit{quantum phase space}, which reduces to the conventional projective Hilbert space in standard quantum mechanics. A decade later, Weinberg established a broad framework for introducing nonlinear corrections to non-relativistic quantum mechanics in a series of seminal papers \cite{weinberg1989precision, weinberg1989particle, weinberg1989testing}. He proposed three experimental approaches to test for nonlinearity: examining spinning particles in external fields, performing Stern-Gerlach-type experiments, and analyzing spectral line broadening.

While this article primarily focuses on secondary nonlinear quantum mechanics in the context of nonlinear optical systems, Weinberg's formulation of nonlinear quantum mechanics will be explored in depth in subsequent sections, as it applies to both primary and secondary types of nonlinearity. Following this, we will examine geometric phases in nonlinear quantum mechanics, which then become a solid substantiation of the main text.

\subsection{Weinberg's formulation of nonlinear quantum mechanics}
In Weinberg's formulation of generalized quantum mechanics, physical states are represented by rays in a complex vector space, while observables correspond to the generating functions of symmetry transformations. Unlike conventional quantum mechanics, where these transformations are strictly linear, Weinberg's framework permits nonlinear symmetry transformations, leading to wave equations that are inherently Hamiltonian in form \cite{weinberg1989precision, weinberg1989testing}.

Specifically, the state of a system is assumed to be described by a complex-valued wave function $\psi$. For a discrete system, the components of the wave functions are denoted by $\psi_k$ of discrete variables $k$ that take values $1, 2, \dots, N$. In general, the wave functions $\psi_k(\boldsymbol{x})$ can also be functions of a position variable $\boldsymbol{x}$. Unlike in conventional quantum mechanics, which describes observables in terms of real bilinear functions $\mathcal{A}(\psi,\psi^*) \equiv A_{kl}\psi_k^*\psi_l$ \cite{dirac1981principles}, the observables in generalized quantum mechanics are represented by real \textit{non-bilinear} functions $\mathcal{O}(\psi,\psi^*)$. Note that not every real non-bilinear function $\mathcal{O}(\psi,\psi^*)$ represents an observable. 

An important feature of conventional quantum mechanics is that the wave functions $\psi$ and $\lambda\psi$ represent the same physical state for arbitrary complex constant $\lambda$. If this feature is retained as a guiding principle to generalized quantum mechanics, those functions representing observables have to obey the homogeneity condition
\begin{equation}
\mathcal{O}(\lambda\psi,\psi^*)=\mathcal{O}(\psi,\lambda\psi^*)=\lambda\mathcal{O}(\psi,\psi^*),
\end{equation}
or equivalently
\begin{equation}\label{Homogeneity}
\psi_k\frac{\partial\mathcal{O}(\psi,\psi^*)}{\partial\psi_k}=\psi_k^*\frac{\partial\mathcal{O}(\psi,\psi^*)}{\partial\psi_k^*}=\mathcal{O}(\psi,\psi^*),
\end{equation}
where the sum of any two observables is defined in the usual way $(\mathcal{O}_a+\mathcal{O}_b)(\psi,\psi^*)\equiv \mathcal{O}_a(\psi,\psi^*)+\mathcal{O}_b(\psi,\psi^*)$, and the multiplication of $\mathcal{O}(\psi,\psi^*)$ by any real number $k$ is another observable $k\mathcal{O}(\psi,\psi^*)$. The product of any two observables is defined by
\begin{equation}\label{MultiplicationLaw}
\mathcal{O}_a(\psi,\psi^*)\cdot\mathcal{O}_b(\psi,\psi^*)\equiv \frac{\partial\mathcal{O}_a(\psi,\psi^*)}{\partial\psi_k}\frac{\partial\mathcal{O}_b(\psi,\psi^*)}{\partial \psi_k^*},
\end{equation}
which is a generalization of the matrix multiplication in conventional quantum mechanics, $\mathcal{A}\cdot\mathcal{B}=A_{kl}B_{lm}\psi^*_k\psi_m$. Under the homogeneity condition \eqref{Homogeneity} and the multiplication law \eqref{MultiplicationLaw}, the norm $n\equiv \psi_k\psi^*_k$ plays the role of a unit element which satisfies $n\cdot\mathcal{O}=\mathcal{O}\cdot n = \mathcal{O}$. As a remark, the product defined by \eqref{MultiplicationLaw} is distributive, but is not necessarily associative and commutative. The non-associative property is the main difference between Weinberg's formalism and conventional quantum mechanics. 

Now we proceed to study the displacement of a state with respect to a symmetry transformation. In conventional quantum mechanics, the displaced states are linear functions of the undisplaced states and thus each displaced state is the result of some linear generators of symmetry applied to the corresponding undisplaced state \cite{dirac1981principles}. If we denote the generator of symmetry by $\mathcal{A}(\psi,\psi^*)=A_{kl}\psi_k^*\psi_l$, the infinitesimal displaced state can be expressed as $\delta\psi_k=-i\epsilon A_{kl}\psi_l$. To generalize this relation, we define the infinitesimal displaced state with respect to a nonlinear function $\mathcal{O}_a(\psi,\psi^*)$ as
\begin{equation}
\delta\psi_k=-i\epsilon\frac{\partial \mathcal{O}_a}{\partial \psi_k^*}.
\end{equation}
Hence, the change in another observable $\mathcal{O}_b(\psi,\psi^*)$ induced by $\mathcal{O}_a(\psi,\psi^*)$ is
\begin{align}
\delta\mathcal{O}_b&=\frac{\partial\mathcal{O}_b}{\partial\psi_k}\delta\psi_k+\frac{\partial\mathcal{O}_b}{\partial\psi_k^*}\delta\psi_k^*\nonumber\\
&=-i\epsilon\left(\frac{\partial\mathcal{O}_b}{\partial\psi_k}\frac{\partial\mathcal{O}_a}{\partial\psi_k^*}-\frac{\partial\mathcal{O}_b}{\partial\psi_k^*}\frac{\partial\mathcal{O}_a}{\partial\psi_k}\right)\nonumber\\
&=i\epsilon(\mathcal{O}_a\cdot\mathcal{O}_b-\mathcal{O}_b\cdot\mathcal{O}_a).
\end{align}
This implies that if we define a commutator by the antisymmetric product, $[\mathcal{O}_a,\mathcal{O}_b]\equiv \mathcal{O}_a\cdot\mathcal{O}_b-\mathcal{O}_b\cdot\mathcal{O}_a$, the change of $\mathcal{O}_b(\psi,\psi^*)$ induced by the symmetry transformation $\mathcal{O}_a(\psi,\psi^*)$ can be compactly expressed as $\delta\mathcal{O}_b = i\epsilon[\mathcal{O}_a,\mathcal{O}_b]$. Obviously, the norm $n=\psi_k^*\psi_k$ is invariant under all symmetric transformations. Even though the product \eqref{MultiplicationLaw} is non-associative, the commutator still satisfies the Jacobi identity
\begin{equation}
[\mathcal{O}_a,[\mathcal{O}_b,\mathcal{O}_c]]+[\mathcal{O}_b,[\mathcal{O}_c,\mathcal{O}_a]]+[\mathcal{O}_c,[\mathcal{O}_a,\mathcal{O}_b]]=0.
\end{equation}
It turns out that the commutators form a Lie algebra of symmetry transformation in a way similar to the case in conventional quantum mechanics, $[\mathcal{O}_i,\mathcal{O}_j]\equiv c_{ij}^k\mathcal{O}_k$, where $c_{ij}^k$ are the structure constants of the Lie algebra.

As a special case, the state $\psi_k(t+\epsilon)$ at a later time, with the process of time displacement generated by a real Hamiltonian function $\mathcal{H}(\psi,\psi^*)$, is determined by the state $\psi_k(t)$ at a earlier time as $\psi_k(t+\epsilon)=\psi_k(t)+\delta\psi_k(t)$, or equivalently
\begin{equation}\label{NonLinearSchrodinerEquation}
i\frac{d\psi_k}{dt}=\frac{\partial \mathcal{H}(\psi,\psi^*)}{\partial \psi_k^*}.
\end{equation}
This is the time-dependent \textit{nonlinear Schr\"{o}dinger equation}. Any observable $\mathcal{O}(\psi,\psi^*)$ of a state $\psi(t)$ and its complex conjugate $\psi^*(t)$ satisfying \eqref{NonLinearSchrodinerEquation} is governed by the equation of motion
\begin{equation}
i\frac{d\mathcal{O}}{dt}=[\mathcal{O},\mathcal{H}],
\end{equation}
as in conventional quantum mechanics. This nonlinear wave equation and its complex conjugate are of the Hamiltonian form, provided that the canonical coordinates and momenta are given by $\psi_k\equiv (q_k+ip_k)/\sqrt{2}$. As a remark, the norm $n$ and the Hamiltonian function $\mathcal{H}$ are invariant under time displacement.

Now we study the class of possible transformations $\psi_k\rightarrow \psi_k'(\psi,\psi^*)$ that leave the equations of motion unchanged. Let $[\mathcal{O}_a,\mathcal{O}_b]_{\psi,\psi^*}$ be the commutator for two observables $\mathcal{O}_a$ and $\mathcal{O}_b$, in which the differentiation is with respect to $\psi_k$ and $\psi_k^*$. A direct calculation gives
\begin{align}
[\mathcal{O}_a,\mathcal{O}_b]_{\psi,\psi^*}&=\frac{\partial \mathcal{O}_a}{\partial \psi_k'}\frac{\partial \mathcal{O}_b}{\partial \psi_l'}[\psi_k',\psi_l']_{\psi,\psi^*}
+\frac{\partial \mathcal{O}_a}{\partial \psi_k'^*}\frac{\partial \mathcal{O}_b}{\partial \psi_l'^*}[\psi_k'^*,\psi_l'^*]_{\psi,\psi^*}\nonumber\\
&=\left(\frac{\partial \mathcal{O}_a}{\partial \psi_k'}\frac{\partial\mathcal{O}_b}{\partial \psi_l'^*}-\frac{\partial \mathcal{O}_b}{\partial \psi_k'}\frac{\partial \mathcal{O}_a}{\partial \psi_l'^*}\right)
[\psi_k',\psi_l'^*]_{\psi,\psi^*}.
\end{align}
Hence $[\mathcal{O}_a,\mathcal{O}_b]_{\psi,\psi^*}=[\mathcal{O}_a,\mathcal{O}_b]_{\psi',\psi'^*}$ if the transformation is canonical, so that
\begin{equation}
[\psi_k',\psi_l']_{\psi,\psi^*}=[\psi_k'^*,\psi_l'^*]_{\psi,\psi^*}=0, [\psi_k',\psi_l'^*]_{\psi,\psi^*}=\delta_{kl}.
\end{equation}
In other words, the commutation relations for any two observables $\mathcal{O}_a$ and $\mathcal{O}_b$ are unchanged under canonical transformation, which implies that the equation of motion $i\dot{\mathcal{O}}=[\mathcal{O},\mathcal{H}]$ is also unchanged under canonical transformation. Hence, the states $\psi_k$ would be transformed by an element of the symplectic group Sp($2N$) \cite{arnol2013mathematical}, as in classical Hamiltonian dynamics, whereas conventional quantum mechanics requires linear transformation, so that only the subgroup U($N$) of Sp($2N$) is used \cite{weyl2016classical}.

A prime example of nonlinear quantum mechanics is a two-component system with just two states $\psi_1$ and $\psi_2$. As will be described in the following section, it is always possible to carry out a canonical transformation so that the Hamiltonian function has the form $\mathcal{H}=n\tilde{\mathcal{H}}(p)$, where $n=|\psi_1|^2+|\psi_2|^2$ and $p\equiv |\psi_2|^2/n$. In conventional quantum mechanics, $\tilde{\mathcal{H}}(p)$ is linear in $p$, but here $\tilde{\mathcal{H}}(p)$ can be a nonlinear function in $p$. A straightforward calculation gives the nonlinear time-dependent Schr\"{o}dinger equation
\begin{subequations}
\begin{align}
i\frac{d\psi_1}{dt}&=[\tilde{\mathcal{H}}(p)-p\tilde{\mathcal{H}}'(p)]\psi_1,\\
i\frac{d\psi_2}{dt}&=[\tilde{\mathcal{H}}(p)+(1-p)\tilde{\mathcal{H}}'(p)]\psi_2.
\end{align}
\end{subequations}
It immediately follows that $p$ is a constant of motion, and the solution of the nonlinear Schr\"{o}dinger equation can simply be written as $\psi_k(t)=c_ke^{-i\omega_k(p)t}$, where $\omega_1(p)\equiv \tilde{\mathcal{H}}(p)-p\tilde{\mathcal{H}}'(p)$ and $\omega_2(p)\equiv \tilde{\mathcal{H}}(p)+(1-p)\tilde{\mathcal{H}}'(p)$. If $\tilde{\mathcal{H}}(p)$ is a linear function in $p$, the frequencies $\omega_1$ and $\omega_2$ are independent of $p$. On the contrary, if $\tilde{\mathcal{H}}(p)$ in a nonlinear function in $p$, the frequencies will depend on the initial conditions, which is expected for nonlinear oscillators.

\subsection{Geometric phases in nonlinear quantum mechanics}
In the previous section, we examined Weinberg’s nonlinear generalization of quantum mechanics. His framework is built on the assumption that the Hamiltonian function $\mathcal{H}(\psi,\psi^*)$ and all observables generating symmetry transformations are homogeneous of degree one in both $\psi$ and $\psi^*$ \cite{weinberg1989precision, weinberg1989testing}. 

Building on this foundation, we now explore geometric phases within Weinberg’s nonlinear quantum mechanics. Specifically, we introduce a canonical transformation of the $N$ components $\psi_k$ of the complex-valued wave function $\psi$ that allows geometric phases to emerge naturally within this framework. This approach parallels the method used by Aharonov and Anandan in conventional quantum mechanics, extending it into the nonlinear domain.

In the following, we consider a nonlinear quantum system, whose $N$ components $\psi_k$ of the wave function $\psi$ evolve according to the nonlinear Schr\"{o}dinger equation $i\dot{\psi}_k = \partial\mathcal{H}(\psi,\psi^*)/\partial\psi_k^*$. Suppose the system undergoes a cyclic evolution so that the initial and final wave functions are related by an overall phase factor, $\psi(T)=e^{i\Theta}\psi(0)$. Then we can construct a projection map which maps an open curve $\psi(t)$ in the Hilbert space to a close curve $\phi(t)$ in the projective Hilbert space via $\psi(t)\equiv e^{if(t)}\phi(t)$, where $f(T)-f(0)= \Theta$. A direct calculation yields
\begin{equation}\label{NonlinearAAPhases}
-\dot{f} = \sum_k\psi_k^*\frac{\partial\mathcal{H}(\psi,\psi^*)}{\partial\psi_k^*}-i\sum_k\phi_k^*\dot{\phi}_k
= \mathcal{H}(\psi,\psi^*)-i\sum_k\phi_k^*\dot{\phi}_k,
\end{equation}
where we have applied the homogeneity condition \eqref{Homogeneity} to the Hamiltonian function $\mathcal{H}(\psi,\psi^*)$ in the last step of \eqref{NonlinearAAPhases}. After integration, the initial and final wave functions are found to be related by
\begin{equation}\label{GeometricPhaseFactor}
\psi(T) = \exp\left\{-i\int_0^T \mathcal{H}(\psi,\psi^*)dt\right\} \exp\left\{-\sum_k\int_0^T\phi_k^*\dot{\phi}_k dt\right\}\psi(0).
\end{equation}
where the first exponential is the dynamical phase factor and the second exponential resembles the geometric phase factor in conventional quantum mechanics. The meaning of this geometric phase will be manifest if we perform a canonical transformation by writing $\psi_k=e^{i\Theta} \sqrt{n} z_k$, or equivalently $\phi_k = \sqrt{n} z_k$, where $n\equiv \sum_k|\psi_k|^2 =\sum_k |\phi_k|^2$. By definition, $z_k$ are subjected to the constraint $\sum_k |z_k|^2=1$. To be specific, we assign the overall phase to be the phase of the $N$-th amplitude, $\Theta\equiv\arg \psi_N$. Thus $z_N$ is real and positive, and can be expressed in terms of other $z_k$ as \cite{weinberg1989precision, weinberg1989testing}
\begin{equation}\label{ProjectiveState}
z_N = \left(1-\sum_{k<N}|z_k|^2\right)^{1/2}.
\end{equation}
Here, the independent dynamical variables are $n$, $\Theta$, and the $z_k$ for $k<N$. These $z_k$ can be expressed directly in terms of the $\psi_k$ as
\begin{equation}
z_k = \frac{\psi_k/\psi_N}{\left(1+\sum_{k<N}|\psi_k/\psi_N|^2\right)^{1/2}}.
\end{equation} 
The advantage of using $z_k$ as the new dynamical variables lies in the fact that the time-dependent nonlinear Schr\"{o}dinger equation has a simpler expression. The homogeneity of the Hamiltonian function allows it to be expressed as $\mathcal{H}(\psi,\psi^*)=n\tilde{\mathcal{H}}(z,z^*)$. Then, the $z_k$ for $k<N$ satisfy the time-dependent nonlinear Schr\"{o}dinger equation $i\dot{z}_k=\partial \tilde{\mathcal{H}}(z,z^*)/\partial z_k^*$. If we define the canonically conjugate variables via $q_k\equiv \sqrt{2}\mbox{Re}\:z_k$ and $p_k\equiv \sqrt{2}\mbox{Im}\:z_k$, these equations are of the Hamiltonian form
\begin{equation}
\frac{dq_k}{dt}=\frac{\partial\tilde{\mathcal{H}}(p,q)}{\partial p_k},
\frac{dp_k}{dt}=-\frac{\partial\tilde{\mathcal{H}}(p,q)}{\partial q_k}.
\end{equation}
A straightforward computation using \eqref{ProjectiveState} yields
\begin{equation}\label{CanonicalTransformationOfGeometricPhase}
\sum_{k=1}^N z_k^*\dot{z}_k = \sum_{k=1}^{N-1}\left( z_k^*\dot{z}_k - \frac{1}{2}\frac{d|z_k|^2}{dt}\right)
=-i\sum_{k=1}^{N-1}\left[p_k\dot{q}_k-\frac{1}{2}\frac{d(p_k q_k)}{dt}\right].
\end{equation} 
Substitution of \eqref{CanonicalTransformationOfGeometricPhase} into \eqref{GeometricPhaseFactor} immediately yields
\begin{equation}\label{ReducedGeometricPhaseFactor}
\psi(T) = \exp\left\{-i\int_0^T n\tilde{\mathcal{H}}(p,q)dt\right\} \exp\left\{i\sum_{k=1}^{N-1}\int_0^T np_k\dot{q}_k dt\right\}\psi(0),
\end{equation}
where we have dropped the integral of the total derivative $\sum_{k} d(p_kq_k)/dt$, since it vanishes identically. Now, the physical meaning of the geometric phase factor $e^{i\gamma} \equiv \exp\left\{i \sum_{k}\int_0^T np_k dq_k\right\}$ is manifest: the geometric phase $\gamma$ is just proportional to the classical actions $I_i\equiv (2\pi)^{-1}\int_0^T p_k dq_k$, where the integrations in this formula must be taken along a path such that the momentum $p_k$ is a single-valued function of the coordinate $q_k$. In other words, the geometric phase can simply be written as $\gamma = 2\pi n\sum_{k<N} I_k$.

As a remark, if the time-dependent wave function exhibits chaotic dynamics in the $2N$-dimensional space of the variables $\mbox{Re}\:z_k$ and $\mbox{Im}\: z_k$, the concept of geometric phase is ill-defined. However, if the system under consideration is a near-integrable Hamiltonian system of the form $\mathcal{H}=\mathcal{H}_0+\mathcal{H}_1$, where $\mathcal{H}_0$ is integrable and $\mathcal{H}_1$ is a small correction to $H_0$, the concept of geometric phase can still be applied to such systems. 

Suppose we have a nonlinear quantum system which depends on a set of external adiabatically varying parameters $\textit{\textbf{R}}$. Then, as we have demonstrated in section \ref{IntegrableSystems}, the new Hamiltonian differs from the angle-independent Hamiltonian $\tilde{\mathcal{H}}_0(I;\textit{\textbf{R}})$ by an amount proportional to the rate of change of $\textit{\textbf{R}}$
\begin{equation}
\tilde{\mathcal{H}}(\theta,I,\textit{\textbf{R}}(t))=\tilde{\mathcal{H}}_0(I;\textit{\textbf{R}}(t))+\frac{d\textit{\textbf{R}}}{dt}\cdot\nabla_{\textit{\textbf{R}}}S(q,I;\textit{\textbf{R}}(t)),
\end{equation}
where $S(q,I;\textit{\textbf{R}}(t))$ is the classical generating function and $\theta_k$ are the angle variables conjugate to the action variables $I_k$. For the case where the external parameters execute periodic motion in the parameter space, one can obtain, in a way similar to the derivation of the Hannay angle, the integration of $\tilde{\mathcal{H}}(\theta,I,\textit{\textbf{R}}(t))$ to be
\begin{equation}\label{ClassicalGeometricPhase}
\int_0^T \tilde{\mathcal{H}}_0(I;\textit{\textbf{R}}) dt + 2\pi \sum_k I_k  - 
\sum_k \oint d\textit{\textbf{R}}\:\langle p_k(\theta,I;\textit{\textbf{R}})\nabla_{\textit{\textbf{R}}}q(\theta,I;\textit{\textbf{R}})\rangle,
\end{equation}
where the bracket stands for the average over the angle variables $\theta_k$ on the torus labelled by $I_k$. Combining \eqref{ReducedGeometricPhaseFactor} and \eqref{ClassicalGeometricPhase}, we find that the initial and final wave functions are related by
\begin{equation}
\psi(T) = \exp\left\{-i\int_0^T n\tilde{\mathcal{H}}_0(I;\textit{\textbf{R}}(t)) dt\right\} e^{i\gamma(I;C)}\psi(0),
\end{equation}
where the geometric phase $\gamma(I;C)$ is evaluated by
\begin{equation}
\gamma(I;C) =n\sum_{k} \int_C d\textit{\textbf{R}}\:\langle p_k(\theta,I;\textit{\textbf{R}})\nabla_{\textit{\textbf{R}}}q(\theta,I;\textit{\textbf{R}})\rangle,
\end{equation}
where $C$ is a closed contour in the parameter space. Using \eqref{ClassicalHannayAngle}, we immediately obtain the connection between Hannay's angles $\Delta\theta_k(I;C)$ and the geometric phase $\gamma(I;C)$
\begin{equation}
\Delta\theta_k(I;C) = -\frac{1}{n}\frac{\partial \gamma(I;C)}{\partial I_k}.
\end{equation}

\section{Derivation of coupled-mode equations in anisotropic fibers with tetragonal symmetry}\label{C}
In this appendix, we describe the evolution of polarization states using the formulation of nonlinear quantum mechanics, and discuss the geometric phases accompanying adiabatic changes in the nonlinear susceptibilities along an anisotropic optical fiber. In the following, we assume the incident radiation to be a nearly monochromatic wave centered at frequency $\omega_0$, so that when light is launched into a nonlinear optical fiber composed of either isotropic or anisotropic media, the electric field can be written as
\begin{equation}
\mathbf{E}(\mathbf{r},t)=\frac{1}{2}\{E_x(\mathbf{r},t)\hat{x}+E_y(\mathbf{r},t)\hat{y}\}e^{-i\omega_0t}+c.c.,
\end{equation}
where $E_x(\mathbf{r},t)$ and $E_y(\mathbf{r},t)$ are the complex amplitudes of the two polarization components of the electric field with carrier frequency $\omega_0$, which are slow-varying function in time. The axial component of the electric field $E_z(\mathbf{r},t)$ is neglected here in accordance with the weakly guiding approximation. In a similar way, the nonlinear part of the induced electric polarization can be written as
\begin{gather}
\mathbf{P}_{NL}(\mathbf{r},t)=\frac{1}{2}\{P_x^{NL}(\mathbf{r},t)\hat{x}+P_y^{NL}(\mathbf{r},t)\hat{y}\}e^{-i\omega_0t}+c.c.,
\end{gather}
where $P_x^{NL}(\mathbf{r},t)$ and $P_y^{NL}(\mathbf{r},t)$ are the complex amplitudes of the nonlinear electric polarization, which are also slow-varying function in time. Employing the same reasoning as in the main text, the slow-varying envelopes of the nonlinear polarization can be written as  \cite{agrawal2000nonlinear}
\begin{align*}
P_i^{NL}(\mathbf{r},t)\approx\frac{\epsilon_0}{4}\sum_{jkl}(\chi^{(3)}_{ijkl}+\chi^{(3)}_{iljk}+\chi^{(3)}_{iklj})E_j(\mathbf{r},t)E_k(\mathbf{r},t)E^*_l(\mathbf{r},t),
\end{align*}
which may also be expressed explicitly as
\begin{align}\label{GeneralNonlinearPolarization}
&P_j^{NL}=\frac{\epsilon_0}{4}\left[3\chi^{(3)}_{jjjj}|E_j|^2E_j+(\chi^{(3)}_{jjkk}+\chi^{(3)}_{jkjk}+\chi^{(3)}_{jkkj})(2|E_k|^2E_j+E_k^2E_j^*)\right.\nonumber\\
&\left.+(\chi^{(3)}_{jjjk}+\chi^{(3)}_{jkjj}+\chi^{(3)}_{jjkj})(2|E_j|^2E_k+E_j^2E_k^*)+3\chi^{(3)}_{jkkk}|E_k|^2E_k\right].
\end{align}
Substituting \eqref{GeneralNonlinearPolarization} into Maxwells's equations, the Fourier transform $\tilde{E}_j(\mathbf{r},\omega-\omega_0)$ of $E_j(\mathbf{r},t)$ is found to obey
\begin{align}
&(\nabla^2 + \epsilon_j k^2) \tilde{E}_j +\frac{k^2}{4}[3\chi^{(3)}_{jjjj}\widetilde{|E_j|^2E_j}\nonumber\\
&+(\chi^{(3)}_{jjkk}+\chi^{(3)}_{jkjk}+\chi^{(3)}_{jkkj})(2\widetilde{|E_k|^2E_j}+\widetilde{E_k^2E_j^*})\nonumber\\
&+(\chi^{(3)}_{jjjk}+\chi^{(3)}_{jkjj}+\chi^{(3)}_{jjkj})(2\widetilde{|E_j|^2E_k}\nonumber\\
&+\widetilde{E_j^2E_k^*})+3\chi^{(3)}_{jkkk}\widetilde{|E_k|^2E_k}]=0,
\end{align}
where $k\equiv \omega/c$ and $\epsilon_j(\omega)\equiv 1+\tilde{\chi}_{jj}^{(1)}(\omega)$ is the principal value of the dielectric function along the $j$ direction. In the weakly guiding limit, a polarization-maintaining fiber supports two non-degenerate polarized components of the HE$_{11}$ mode. Similar to the reasoning applied to isotropic materials, we may assume that the guided modes supported by the fiber are not significantly distorted by nonlinear effects, so that the nonlinear part of the dielectric function can be treated as a small perturbation to its linear counterpart. Based upon these approximations, the slow-varying functions of the electric field are written as
\begin{equation}
E_j(\mathbf{r},t)=A_j(z,t)F_j(x,y)\exp(i\beta_{0j}z),\:\;\mbox{where}\:\;j=x\:\;\mbox{or}\;\:y,
\end{equation}
where $A_j(z,t)$ are the slow-varying wave amplitudes, $F_j(x,y)$ are the transverse distribution of the two polarization components of the fundamental HE$_{11}$ mode, and $\beta_{0j}$ are the modal propagation constants in the absence of dispersion and nonlinearity. Here we require that the transverse modal distribution $F_j(x,y)$ obey $(\nabla_t^2+n_j^2k^2-\beta_j^2)F_j=0$, where $n_j^2\equiv 1+\mbox{Re}\tilde{\chi}_{jj}^{(1)}(\omega)$.  After employing the same method as for deriving coupled-mode equations for isotropic materials, the slow-varying wave amplitudes $A_j(z,t)$ are found to obey the following set of coupled-mode equations  \cite{agrawal2000nonlinear}
\begin{align}\label{AnisotropicNonlinearSchrodingerEquation}
&i\left(\frac{\partial A_j}{\partial z}+\beta_{1j}\frac{\partial A_j}{\partial t}+\frac{\alpha_j}{2}A_j\right)-\beta_{2j}\frac{\partial^2A_j}{\partial t^2}+a_j|A_j|^2A_j\nonumber\\
&+b_j\left(2|A_k|^2A_j+A_k^2A_j^*e^{2i(\beta_{0k}-\beta_{0j})z}\right)+c_j\left(2|A_j|^2A_k e^{i(\beta_{0k}-\beta_{0j})z}\right.\nonumber\\
&\left.+A_j^2A_k^*e^{i(\beta_{0j}-\beta_{0k})z}\right)+d_j|A_k|^2A_ke^{i(\beta_{0k}-\beta_{0j})z}=0,
\end{align}
where $\alpha_j\equiv \frac{k_0^2}{2\beta_{0j}}\mbox{Im}\tilde{\chi}_{jj}^{(1)}(\omega_0)$ accounts for fiber losses, and $\beta_{1j}$ and $\beta_{2j}$ are the first and second order derivatives of $\beta_j$ evaluated at the carrier frequency $\omega_0$ respectively.  $\beta_{1j}=v_{gj}^{-1}$ is the inverse of the modal group velocity, and any difference between $\beta_{1x}$ and $\beta_{1y}$ leads to modal dispersion effects. $\beta_{2j}=\partial v_{gj}^{-1}/\partial \omega$ is the modal group velocity dispersion parameter, which is responsible for dispersive pulse broadening for the $j$-th polarization component. The remaining parameters $a_j$, $b_j$, $c_j$ and $d_j$ are defined by
\begin{align}\label{gamma}
&a_j\equiv \gamma_{jjjj}, b_j\equiv\frac{1}{3}(\gamma_{jjkk}+\gamma_{jkjk}+\gamma_{jkkj}),\nonumber\\
&c_j\equiv\frac{1}{3} (\gamma_{jjkj}+\gamma_{jkjj}+\gamma_{jjjk}), d_j\equiv \gamma_{jkkk},
\end{align}
where $\gamma_{jklm}$ are the nonlinearity parameters defined by
\begin{equation}\label{NonlinearAnisotropicParameters}
\gamma_{jklm}\equiv\frac{3k_0^2}{8\beta_0}\chi_{jklm}^{(3)}\int_{-\infty}^{\infty}\int_{-\infty}^{\infty}|F|^4 dxdy \Big/\int_{-\infty}^{\infty}\int_{-\infty}^{\infty}|F|^2 dxdy. 
\end{equation}
Here we assume that the difference between the two propagation constants, $\beta_{0x}-\beta_{0y}$, is a small quantity when compared to the values of $\beta_{0x}$ and $\beta_{0y}$, so that we may apply the approximations $\beta_{0x}\approx\beta_{0y}\approx\beta_0$ and $F_x\approx F_y \approx F$ when evaluating the integral \eqref{NonlinearAnisotropicParameters}. 

The terms involving exponential factors in \eqref{AnisotropicNonlinearSchrodingerEquation} are due to coherent couplings between the two polarization components, and their importance are determined by the strength of the modal birefringence $B_m\equiv \Delta\beta/k_0$ or the beat length $L_B\equiv 2\pi/\Delta\beta$. If the fiber length $L$ is much longer than the beat length $L_B$, the oscillating terms in \eqref{AnisotropicNonlinearSchrodingerEquation} are fast-oscillating function of $z$ and may be neglected. For low-birefringent fibers with a typical beat length $L_B\approx 1$ m, the oscillating terms in \eqref{AnisotropicNonlinearSchrodingerEquation} should be retained, especially for fibers with short lengths from a few percents to a few tenths of the beat length \cite{kuzin2001measurements, ibarra2003measurement}.

We now discuss anisotropic fibers with tetragonal symmetry, which are used in the main text. The tetragonal crystal system has seven point groups, $4$, $\bar{4}$, $4/m$, $422$, $4mm$, $\bar{4}2m$ and $4/mmm$ \cite{bradley2010mathematical}. For the three crystal classes $4$, $\bar{4}$ and $4/m$, the third-order nonlinear susceptibility tensor $\chi^{(3)}$ has sixteen nonzero elements, of which only eight are independent, and its elements obey \cite{boyd2003nonlinear}.
\begin{equation}\label{chi}
  \begin{array}{l}
  \left\{
   \begin{array}{c}
   \chi^{(3)}_{xxyy} = \chi^{(3)}_{yyxx},\\
   \chi^{(3)}_{xyxy} = \chi^{(3)}_{yxyx},\\
   \chi^{(3)}_{xyyx} = \chi^{(3)}_{yxxy},\\
   \chi^{(3)}_{xxxx}=\chi^{(3)}_{yyyy},
   \end{array}
  \right.
  \left\{
   \begin{array}{c}
   \chi^{(3)}_{yyxy} = -\chi^{(3)}_{xxyx},  \\
   \chi^{(3)}_{yxyy} = -\chi^{(3)}_{xyxx},  \\
   \chi^{(3)}_{xyyy} = -\chi^{(3)}_{yxxx},  \\
   \chi^{(3)}_{xxxy}=-\chi^{(3)}_{yyyx}.
   \end{array}
  \right.  \\ 
  \end{array}
\end{equation}
Hence, from the above symmetry requirements, the parameters $a_j$, $b_j$, $c_j$ and $d_j$ in \eqref{gamma} satisfy
\begin{equation}
  \begin{array}{l}
  \left\{
   \begin{array}{c}
   a_x = a_y = a,\\
   b_x = b_y = b,
   \end{array}
  \right.
  \left\{
   \begin{array}{c}
   c_x = -c_y = c,  \\
   d_x = -d_y = d,
   \end{array}
  \right.  \\ 
 \end{array}
\end{equation}
so that $u_j$ obeys the following coupled-mode equations in the retarded frame
\begin{align}\label{4bar44m}
i\frac{\partial u_j}{\partial z}=&-\frac{\xi_j}{2}u_j+\frac{\beta_2}{2}\frac{\partial^2 u_j}{\partial \tau^2}-a|u_j|^2u_j-b\left(2|u_k|^2u_j+u_k^2u_j^*\right)\nonumber\\
&\mp c\left(2|u_j|^2u_k+u_j^2u_k^*\right)\mp d|u_k|^2u_k,
\end{align}
where the minus and plus signs correspond to $j=x$ and $j=y$ respectively. \eqref{4bar44m} is integrable only when the condition $d=-c$ is fulfilled. This condition leads to
\begin{align}\label{IntegrableNSE}
i\frac{\partial u_j}{\partial z}=&-\frac{\xi_j}{2}u_j+\frac{\beta_2}{2}\frac{\partial^2 u_j}{\partial \tau^2}-a|u_j|^2u_j-b\left(2|u_k|^2u_j+u_k^2u_j^*\right)\nonumber\\
&\mp c\left(2|u_j|^2u_k+u_j^2u_k^*-|u_k|^2u_k\right).
\end{align}
For fibers with negligible losses, \eqref{IntegrableNSE} can be written in the form of the nonlinear Schr\"{o}dinger equation, and the Hamiltonian is given by
\begin{align}
H&=-\frac{1}{2}\int d\tau\left\{\beta_2\left(\left|\partial_\tau u_x\right|^2+\left|\partial_\tau u_y\right|^2\right)\right.\nonumber\\
&\left.+\Delta\beta (|u_x|^2-|u_y|^2)+c_0(|u_x|^2+|u_y|^2)^2\right.\nonumber\\
&\left.+c_z(|u_x|^2-|u_y|^2)^2+c_x(u_x^*u_y+u_y^*u_x)^2\right.\nonumber\\
&\left.+2c(|u_x|^2-|u_y|^2)(u_x^*u_y+u_y^*u_x)\right\},
\end{align}
which may also be expressed as
\begin{align}\label{TetragonalHamiltonian}
H=&-\frac{1}{2}\int d\tau\left\{\beta_2\left(\left|\partial_\tau u_x\right|^2+\left|\partial_\tau u_y\right|^2\right)+\Delta\beta S_z\right.\nonumber\\
&\left.+c_0S_0^2+c_zS_z^2+c_xS_x^2+2cS_zS_x\right\},
\end{align}
where $c_0\equiv \frac{1}{2}(a+b)$, $c_z\equiv \frac{1}{2}(a-b)$ and $c_x\equiv b$. For the four crystal classes $422$, $4mm$, $\bar{4}2m$ and $4/mmm$, the third-order nonlinear susceptibility tensor $\chi^{(3)}$ has eight nonzero elements and only four of them are independent. The nonzero elements of the third-order nonlinear susceptibility obey \eqref{chi}, which leads to a set of coupled-mode equations in the form of \eqref{IntegrableNSE}.

\section{General expression of homogeneous Hamiltonian}\label{D}

In this appendix, we present a reasoning of why only specific coupled-mode equations can be written in the form of nonlinear Schr\"{o}dinger equations. Let us consider a Hamiltonian $H(u_x,u_y)$ that is homogeneous in $u_x$ and $u_y$, which obey
\begin{equation}\label{homogeneous}
H(\lambda u_x,\lambda u_y)=H(u_x,u_y),
\end{equation}
where $\lambda$ is an arbitrary unit complex number. To simplify the discussion, we assume that $u_x$ and $u_y$ are functions of only $z$. In order to fulfill the condition \eqref{homogeneous}, the Hamiltonian must be a combination of $|u_x|^2$, $|u_y|^2$, $u_xu_y^*$, and $u_yu_x^*$. Hence, up to quartic terms, we may write the Hamiltonian in the form $H=H^{(2)}+H^{(4)}$, where
\begin{subequations}
\begin{align}\label{Hamiltonian2}
H^{(2)}&=A|u_x|^2+B|u_y|^2+Cu_xu_y^*+Du_yu_x^*,\\
H^{(4)}&=E_1|u_x|^4+E_2|u_y|^4+F_1|u_x|^2u_xu_y^*+F_2|u_y|^2u_yu_x^*\nonumber\\
&+G_1|u_x|^2u_yu_x^*+G_2|u_y|^2u_xu_y^*+H|u_x|^2|u_y|^2\nonumber\\
&+I_1u_x^2u_y^{*2}+I_2u_y^2u_x^{*2}\label{Hamiltonian4}
\end{align}
\end{subequations}
As the Hamiltonian must be a real function of $u_x$ and $u_y$, $H^*(u_x,u_y)=H(u_x,u_y)$, we have further restrictions on the coefficients: both $A$, $B$, $E_1$, $E_2$, and $H$ are real numbers, where the remaining coefficients satisfy
\begin{equation}\label{restrictions}
D=C^*,G_1=F_1^*,G_2=F_2^*,I_2=I_1^*.
\end{equation}
Now, if we assume the coupled-mode equations can be written in the form of nonlinear Schr\"{o}dinger equations, \eqref{Hamiltonian2} - \eqref{Hamiltonian4}, and \eqref{restrictions} immediately yield
\begin{subequations}
\begin{align}\label{GeneralNonlinearEquation1}
i\dot{u}_x&=\frac{\partial H}{\partial u_x^*}=Au_x+C^*u_y+2E_1|u_x|^2u_x+F_1u_x^2u_y^*+F_2|u_y|^2u_y\nonumber\\
&+2F_1^*|u_x|^2u_y+H|u_y|^2u_x+2I_1^*u_y^2u_x^*,\\
i\dot{u}_y&=\frac{\partial H}{\partial u_y^*}=Bu_y+Cu_x+2E_2|u_y|^2u_y+F_1|u_x|^2u_x+F_2u_y^2u_x^*\nonumber\\
&+2F_2^*|u_y|^2u_x+H|u_x|^2u_y+2I_1u_x^2u_y^*.\label{GeneralNonlinearEquation2}
\end{align}
\end{subequations}
As we can see in the primary article, the general form of coupled-mode equations when $u_x$ and $u_y$ are functions of only $z$ is
\begin{align}\label{GeneralCoupledModeEquations}
&i\frac{d u_j}{d z}+\frac{\xi_j}{2}u_j+a_j|u_j|^2u_j+b_j\left(2|u_k|^2u_j+u_k^2u_j^*\right)\nonumber\\
&+c_j\left(2|u_j|^2u_k+u_j^2u_k^*\right)+d_j|u_k|^2u_k=0,
\end{align}
Comparing \eqref{GeneralNonlinearEquation1} - \eqref{GeneralNonlinearEquation2} and \eqref{GeneralCoupledModeEquations}, we have $\xi_x=-2A$, $\xi_y=-2B$,
\begin{equation}
  \begin{array}{l}
  \left\{
   \begin{array}{c}
   a_x=-2E_1,\\
   a_y=-2E_2,\\
   d_x=-F_2,\\
   d_y=-F_1,
   \end{array}
  \right.
  \left\{
   \begin{array}{c}
   b_x=-H/2=-2I_1^*,  \\
   b_y=-H/2=-2I_1,  \\
   c_x=-F_1^*=-F_1,  \\
   c_y=-F_2^*=-F_2.
   \end{array}
  \right.  \\ 
  \end{array}
\end{equation}
which implies that $F_1$, $F_2$, and $I_1$ are real numbers, and hence both $\xi_j$, $a_j$, $b_j$, $c_j$, and $d_j$ are real numbers, and obey
\begin{equation}\label{GeneralRestrictions}
b_x=b_y=b,d_y=c_x,d_x=c_y.
\end{equation}
As long as the condition \eqref{GeneralRestrictions} is fulfilled, the coupled-mode equations can be written in the form of nonlinear Schr\"{o}dinger equations, and the associated Hamiltonian has the form
\begin{align*}
&H=-\left[\frac{\xi_x}{2}|u_x|^2+\frac{\xi_y}{2}|u_y|^2+\frac{a_x}{2}|u_x|^4+\frac{a_y}{2}|u_y|^4+2b|u_x|^2|u_y|^2\right.\nonumber\\
&\left.+(c_x|u_x|^2+c_y|u_y|^2)(u_xu_y^*+u_yu_x^*)+\frac{b}{2}(u_x^2u_y^{*2}+u_y^2u_x^{*2})\right].
\end{align*}

\end{appendix}

\bibliography{references}

\end{document}